\documentclass[aps,prd,twocolumn,amsmath,groupedaddress]{revtex4}

\usepackage{graphicx}
\usepackage{bm}
\usepackage{dcolumn}

\newcommand{\be}{\begin{eqnarray}}     	\newcommand{\ee}{\end{eqnarray}}   
\newcommand{\ba}{\begin{array}}         \newcommand{\ea}{\end{array}} 
\newcommand{\bs}{\begin{subequations}}  \newcommand{\es}{\end{subequations}} 
\newcommand{\rf}[1]{~(\ref{#1})}  \newcommand{\rfd}[2]{~(\ref{#1},\,\ref{#2})}
\newcommand{\rfs}[2]{~(\ref{#1}--\ref{#2})}
\newcommand{\ct}[1]{\ Ref.~\cite{#1}}  \newcommand{\cts}[1]{\ Refs.~\cite{#1}}
\newcommand{\lb}[1]{\label{#1}}         
\newcommand{\nn}{\nonumber}
\newcommand{\rav}{\,{=}\,}              \newcommand{\sh}[1]{\,{#1}\,}
\newcommand{\lf}{\left}                 \newcommand{\rt}{\right}
\newcommand{\fr}{\frac}

\def\pd{\partial}  \def\d{\delta}    \def\D{\Delta}
\def\phr{\phi_r}   \def\phc{\phi_c}

\def\t{\tau}       \def\b{\beta}     \def\Nb{\nabla}	
\def\tt{\tau_c}
\def\s{\sigma}     \def\eps{\epsilon} \def\g{\gamma} 
\def\n{\^{\bf n}}
\def\r{{\bf r}} 	\def\k{{\bf k}} 
\def\^{\hat}	        \def\~{\tilde}

\newcommand{\sign}{\mathop{\rm sign}}
\newcommand{\ci}{\mathop{\rm ci}}     

\def\vp{u}
\def\dd{\d_{\rm D}}

\begin{document}


\title{Dynamics of Cosmological Perturbations\\
               in Position Space}


\author{Sergei Bashinsky}
\affiliation{Department of Physics \\
              Princeton University \\
           Princeton, New Jersey 08544}

\author{Edmund Bertschinger}
\affiliation{Department of Physics \\
       Massachusetts Institute of Technology \\
           Cambridge, Massachusetts 02139\\
            {\rm(May 8, 2002)}}

\begin{abstract}

  We show that the linear dynamics of cosmological perturbations 
can be described by coupled wave equations, allowing their efficient 
numerical and, in certain limits, analytical integration directly 
in position space.  The linear evolution of any perturbation can 
then be analyzed with the Green's function method.  Prior to hydrogen 
recombination, assuming tight coupling between photons and baryons, 
neglecting neutrino perturbations, and taking isentropic (adiabatic) 
initial conditions, the obtained Green's functions for all metric, 
density, and velocity perturbations vanish beyond the acoustic horizon.  
A localized primordial cosmological perturbation expands as an acoustic 
wave of photon-baryon density perturbation with narrow spikes at its 
acoustic wavefronts.  These spikes provide one of the main contributions 
to the cosmic microwave background radiation anisotropy on all experimentally
accessible scales.  The gravitational interaction between cold dark matter 
and baryons causes a dip in the observed temperature of the radiation
at the center of the initial perturbation.  We first model the radiation 
by a perfect fluid and then extend our analysis to account for finite 
photon mean free path.  The resulting diffusive corrections smear the 
sharp features in the photon and baryon density Green's functions over 
the scale of Silk damping.

\end{abstract}


\maketitle


\section{Introduction}
\label{sec_int} 

The nearly perfect black body spectrum and isotropy of the cosmic 
microwave background radiation indicates that the universe at
large redshift was highly uniform and in thermal equilibrium,
with fluctuations in temperature of only about 1 part in $10^5$ on 
observable scales.  
Under these conditions, the dynamics of matter, radiation,
and gravity is described accurately by linearizing the governing
equations about their spatially homogeneous solutions 
representing an unperturbed expanding background.
In place of the strongly nonlinear fluid and Einstein
equations, we have a system of coupled linear partial differential
equations.  The linear approximation continues to apply much
later on large scales even after nonlinear structures such as stars
and galaxies have formed on smaller scales.

The cosmological perturbations can be described by a set of 
classical fields, {\it e.g.\/} the gravitational potential~$\phi$.
After the perturbations are created in the early universe,
each field may be expanded over a convenient set of basis 
functions 
\be
   \phi(\r,\tau_i)=\sum_k\phi_k f_k(\r)~,
\lb{phi0}
\ee
where the $\phi_k$ are expansion coefficients
at some initial time~$\t_i$.
Under linear evolution at a later time~$\t$
\be
   \phi(\r,\tau)=\sum_k\phi_kF_k(\r,\tau)\ ,
\lb{phievol}
\ee
where $F_k(\r,\tau)$ is the solution to the linearized 
equation for~$\phi$ that satisfies
the initial conditions $F_k(\r,\tau_i)=f_k(\r)$.

The range of possible basis functions~$f_k(\r)$ is enormous.  
Since the pioneering work by Lifshitz,\ct{Lifshitz46},
nearly all the cosmological perturbation theory calculations, {\it
e.g.\/}\cts{Bard80,KS84,Mukh,LiddleLyth93,MaBert,HuSug96},
have used harmonic plane waves or their generalizations
in curved spaces.  There is a good reason for this: because the
dynamical equations are translationally invariant, each harmonic mode evolves
unchanged aside from a time-dependent multiplicative factor
$T_k(\tau)$ called the transfer function: $F_k(\r,\tau)=T_k(\tau)\,f_k(\r)$.
This separation of variables allows the partial differential equations
to be reduced to a set of ordinary differential equations
that are straightforward to solve numerically.

However attractive this reduction appears, it by no means implies
that cosmological perturbation theory reduces to the simplest form
in Fourier space.
The plane wave expansion may be thought of as a localized
(Dirac delta) basis in Fourier space.  From this perspective, it is
not unreasonable to consider a localized basis in real 
space.
In this case, the function $F_k(\r,\tau)$ in eq.\rf{phievol} 
no longer factors into a simple product.  The lack
of separation of variables would seem to imply that perturbation
theory is more difficult in position space.  In fact, we will show
that it is not only simple in the linear regime, but
the dynamics is clearer and more intuitive in real space.

Suppose, for example, that we consider the evolution of 
a perturbation originating from a point disturbance
\be
   \phi(\r,\t_i\to 0)=f_k(\r)=\delta_{\rm D}^{(3)}(\r-\r_0)\ ,
\lb{fdelta}
\ee
where $\delta_{\rm D}^{(3)}(\r-\r_0)=\delta_{\rm D}(x-x_0)
\delta_{\rm D}(y-y_0)\delta_{\rm D}(z-z_0)$
is the product of Dirac delta functions.  
What will such an initial perturbation evolve to by time~$\tau$?  
The corresponding function $F_k(\r,\tau)=G(\r-\r_0,\tau)$ is called 
a Green's function.
To see how the Green's function language can be simpler than the 
transfer function approach, we notice that,
as shown in\ct{BB_let}, prior to recombination the primordial
isentropic (adiabatic) perturbation satisfying the initial 
conditions\rf{fdelta}
expands in the photon-baryon plasma as a spherical acoustic
wave with sound speed $c_s$.  During the radiation era, $c_s=
c/\sqrt{3}$ and the gravitational potential is
simply given by the properly normalized Heaviside step function:
\be
   \phi(r,\tau)={3\over4\pi(c_s\tau)^3}
\lf\{
\ba{cl}
   1~, & r\le c_s\tau~, \\
   0~, & r>c_s\tau~.
\ea
\rt. 
\lb{radwave}
\ee
Here, $r$ is the comoving distance from the initial
point perturbation, and  
$\t$ is conformal time related to the proper time~$t$ 
by $d\t=dt/a(t)$, where $a(t)$ is the cosmological scale factor.
Later we will also use Green's functions produced by 
initial disturbances on two-dimensional planes in space.

The position space representation is formally equivalent to the
Fourier space representation.  It can be used to describe the
dynamics of cosmological perturbations regardless of their
origin and statistical properties.  
The Green's function method can be 
applied just as well to the evolution of 
nearly scale-invariant perturbations 
generated by inflation,\cts{BB_let,SB_Thesis}, 
as it can be applied to local perturbations produced by topological 
defects,\cts{VeerSteb90,Magueijo92}.

So what new can be learned from
this approach?  Green's functions contain the same information
as transfer functions, but that information is packaged differently.
The Green's function approach reveals new sides to cosmological
dynamics that are of both phenomenological and theoretical 
interest.

Phenomenologically, Green's functions are often characterized
by localized features such as the acoustic wavefront.  Through the 
uncertainty relation $\Delta x \Delta k \gtrsim1$,
localization in position space results in features being spread over a broad
range of wavenumbers in Fourier space.  An example
of this is the acoustic peaks in the cosmic microwave background (CMB)
power spectrum $C_l$.  In\ct{BB_let} we have shown that the 
position-space analogue of $C_l$, the angular correlation
function $C(\theta)$, has localized features instead of acoustic
oscillations.  These features offer an alternative signature for
experimentalists to measure.

Theoretically, acoustic and transfer processes are made explicit
in position space.  This offers new methods for solving the
evolution equations or leads to substantially simpler equations
and solutions in certain cases.  An example is the spherical
wave solution for the radiation era given by eq.\rf{radwave}.  
The acoustic wave is manifest, and this suggests that the position 
space view may provide a more direct understanding of the 
dependence of CMB anisotropy patterns on the 
underlying cosmological parameters.

Equations of perturbation dynamics can in principle  
be numerically solved faster in position space than the equivalent equations 
in Fourier space for same desired accuracy. 
This is because the Green's functions are monotonic and
limited in their spatial extent by the acoustic horizon for
perfect fluids, while the Fourier transfer
functions oscillate in both the wavenumber and time coordinates
requiring a larger number of sample points for their accurate 
representation.

This paper describes cosmological linear perturbation theory
in position space using the Green's function approach.  
The discussion is primarily focused on the period of cosmological
evolution prior to hydrogen recombination and radiation decoupling 
from baryonic matter.  In the current paper we derive evolution 
equations that are convenient for a position space analysis
and consider their Green's function solutions. 
In a later paper we will show how these results may be used to 
describe CMB anisotropy.  Sec.~\ref{sec_frm} presents 
the dynamical equations describing coupled perturbations in the 
metric and in radiation and matter 
modeled by locally isotropic fluids.
In Sec.~\ref{sec_inc} we specify the initial conditions 
for the Green's functions and give explicit Green's function  
solutions in the radiation epoch.
In Sec.~\ref{sec_res} we discuss general properties
of the Green's functions and their numerical integration in
the fluid model, and we analyze the results of numerical 
integration up to the time of hydrogen recombination.
Sec.~\ref{sec_gen} shows how the position-space description
of perturbations can be implemented when the fluid equations of 
Sec.~\ref{sec_frm} are replaced by the full equations of Boltzmann phase 
space dynamics,
which are summarized in the Appendix.  Then in Sec.~\ref{sec_slk}
we find the leading corrections to the fluid approximation
results.  A brief summary is given in the Conclusion. 

\begin{table*}[t]
\begin{tabular}{|c|l|c|}
\hline
Symbol  & \qquad\qquad\qquad\qquad  Meaning  &    Defining Equation \\
\hline

$a$       &   Scale factor relative to the present  &   \rf{pertrw}     \\
$y$       &   Scale factor relative to the radiation-matter density equality
                                                        & \rf{ybta_def} \\  
$\t$      &   Conformal time      &   
                                      \rf{pertrw}     \\
$\t_e$    &   Characteristic $\t$
          of the radiation-matter density equality  &   \rf{te_def}   \\
$\r$      &   Comoving $3$-space coordinate     &   \rf{pertrw}   \\ 

superscript $(3)$ &  Spherical (3D) Green's function & \rf{ic3}   \\
superscript $(1)$ & Plane-parallel (1D) Green's function &   \rf{ic1}   \\
$x$       &   Spatial coordinate of a plane-parallel Green's
                                               function  &   \rf{ic1}   \\
superscript $($T$)$ & Fourier space transfer function    &   \rf{trf_def}   \\
$\k$      &   Comoving wave vector           &   \rf{FTdef}   \\
$k$       &   $|\k|$              &    --  \\
$\phi,\psi$   &   Metric perturbations  &        \rf{pertrw}  \\
subscript $\gamma$  &  Photons        &    --     \\
subscript $b$   &   Baryons           &    --     \\
subscript $r$   &   Radiation fluid (coupled photons and baryons) &  --    \\
subscript $d$   & Difference in photon and baryon perturbations &    --    \\
subscript $\nu$ & Massless neutrinos  &    --     \\
subscript $c$   &   Cold dark matter (CDM) &    --     \\
subscript rel &  Background of relativistic species ($\g$ plus $\nu$) 
                                                            &    --     \\
subscript $\rm m$ & Background of non-relativistic matter (baryons plus CDM) 
                                                            &    --     \\
$\d$      &    Energy density enhancement / &     \rf{T00}   \\ 
          &    If applied to a variable, the perturbation 
               of that variable             &        --      \\ 
$\dd(x)$  &    Dirac delta function         &        --      \\ 
$\vp$     &    Peculiar velocity potential  &     \rf{T0i}   \\
$v_i$     &    Peculiar 3-velocity          &     \rf{T0i}   \\
$\d p$    &    Pressure perturbation        &    ~(\ref{T3},\,\ref{pi_def}) \\
$\pi$     &    Shear stress potential       &     \rf{pi_def}   \\
$\sigma$  &    Entropy perturbation potential & 
                                   ~(\ref{eta_def}--\ref{sigma_def}) \\
$f$       &  Phase space distribution              & \rf{f_def}  \\  
$F$       &  Energy averaged perturbation of $f$   & \rf{Fdef}   \\
$f_l$     &  Potentials for angular moments of $F$ & \rf{fl_def} \\  
$\b$      &  Baryonic fraction of nonrelativistic matter 
                         ($\rho_b/\rho_{\rm m}$)   & \rf{ybta_def} \\
$c_s$     &  Speed of sound in the photon-baryon plasma & 
                                   ~(\ref{c_s_def},\,\ref{c_s_exp}) \\
$c_w$     &  ``Isentropic sound speed'' $\lf(\pd p /
                     \pd\rho\rt)^{1/2}_{\rm adiab}$  
                                                  & \rf{c_w_def} \\
$A$ and $B$ & $1/(3c_w^2)$ and $1/(3c_s^2)$ respectively  & \rf{AB_def}  \\  
$S(\t)$       &  Radius of acoustic horizon           & \rf{S_def}   \\      
$\tt$     &  Mean conformal time of a photon collisionless flight   
                                                  & \rf{tau_c_def}   \\      
\hline
\end{tabular}
\caption{Frequently used notations.}
\lb{tab_notations}
\end{table*}
We end this introduction with a summary of our
conventions and notations.  Throughout the paper, Greek
indices range from 0 to 3 and label components of space-time
vectors.  Components of spatial 3-vectors carry Latin indices
ranging from 1 to 3; if the indices are omitted,
3-vectors are typed in bold.  The speed of light is $c=1$.
 The $2\pi$ factors in the Fourier transforms
always appear in the denominator of the momentum integral,
as in the equation
\be
\phi(\r)=\int \fr{d^3k}{(2\pi)^3}\,e^{i\k\cdot\r}\phi(\k)~.
\lb{FTdef}
\ee
We consider only the scalar perturbation mode, the one involving
radiation and matter density perturbations, and work in the conformal 
Newtonian gauge for the perturbed Robertson-Walker 
metric,\cts{MaBert,Mukh}.  In this gauge
\be
   ds^2=a^2(\t)\left[-(1+2\phi)d\tau^2+(1-2\psi)d\r^2\right]\ ,
\lb{pertrw}
\ee
where $d\r^2$ is the three-metric of a Robertson-Walker space.
For perfect fluids, there is a single gravitational 
potential $\phi=\psi$, but in general two distinct potentials are required.
(Note that the perturbation $\phi$ of the metric\rf{pertrw}
is called $\psi$, and $\psi$ is $\phi$, in\cts{MaBert, Bert_LesHouches}.
Our present choice agrees with\ct{Mukh}.)
Other frequently used variables and notations are summarized
in Table 1.  They will be introduced systematically in what follows.

\section{Cosmological Dynamics in the Fluid Approximation}
\label{sec_frm} 

%
%
\subsection{The model}
\lb{subsec_model}

In this and the following two sections we study perturbation
dynamics adopting a simplified model where the photon-baryon plasma and cold 
dark matter are approximated by coupled fluids.  The photon gas is assumed 
to behave as a locally isotropic fluid characterized by its density 
and velocity at every point in space and its pressure equals one
third of the photon energy density.  This is a good approximation before
recombination when photons intensively scatter against free electrons
and, because of the Coulomb interaction between the electrons and baryons, 
the velocity of baryons is locked to equal the mean velocity of 
the photon fluid.  We will arrive at these results consistently from
the Boltzmann equation and will consider the leading corrections to 
the fluid approximation in Secs.~\ref{sec_gen}--\ref{sec_slk}.
The effects of global curvature and of the cosmological constant
should be very small on the scale of the acoustic dynamics
before recombination and we do not include them
in the discussion.

Neutrinos contribute a large fraction 
(about $40\%$)
of the radiation energy and require specification of their 
full phase space distribution 
even before recombination.  We do not include a full treatment 
of neutrino perturbations
in the present paper for the following reasons:
The fluid model, corrected for photon diffusion
and for the neutrino contribution to the background energy density
as shown below, can describe perturbations in the early universe 
at least up to $5\%$ 
accuracy,\ct{SB_Thesis}, when compared in its predictions of CMB anisotropy 
with full numerical calculations.
The fluid description is adequate for baryons, dark matter,
and photons before recombination. Its simplicity and intuitive
appeal make the fluid approximation an attractive starting
point for applying perturbation theory in position space. 
The position space approach can also be used
for accurate description of neutrino dynamics and offers 
substantial advantages over the traditional Fourier 
decomposition, but that requires new constructions different 
from the spirit of the fluid description,\ct{SB_Thesis}. 
We postpone the phase space analysis of neutrino perturbations
to a later paper.
 
To account for the substantial contribution of neutrinos to 
the background radiation density while excluding them 
cleanly from the perturbation dynamics, we consider a fictitious 
universe filled with photons, baryons, and cold dark matter~(CDM),
but without neutrinos. In this model universe 
the photon background energy density,~$\rho_{\g}^{\rm(model)}$,
equals the actual (physical) energy density of the combined relativistic
backgrounds of photons and neutrinos:
\be
\rho_{\g}^{\rm(model)}
  = \rho_{\rm rel}^{\rm(phys)}
  \equiv \rho_{\g}^{\rm(phys)}+\rho_{\nu}^{\rm(phys)}
  = \lf(1+R_{\nu}\rt)\rho_{\g}^{\rm(phys)}
\nn
\ee
where
\be
R_{\nu}\equiv \frac{\rho_{\nu}}{\rho_{\g}+\rho_{\nu}}
         = \lf[1+\fr{8}{7N_{\nu}^{\rm eff}}\lf(\fr{11}{4}\rt)^{4/3}\rt]^{-1}~,
\nn
\ee
is the fraction of the radiation energy density in neutrinos
that equals $R_{\nu}\simeq 0.408$ for $N_{\nu}^{\rm eff}\simeq 3.04$,
{\it e.g.} from\ct{N_nu_value}.

The sound speed in the photon-baryon fluid is determined by 
the ratio of baryon and photon energy densities as 
\be
c_s^2\equiv \left(\frac{\pd p_{\gamma}}{\pd(\rho_{\gamma}+\rho_{\rm b})}
       \right)_{\rm adiab}=\fr1{3\lf[1+(3\rho_b)/(4\rho_{\g})\rt]}~.
\lb{c_s_def}
\ee
The sound speed controls the dynamics of acoustic perturbations
via the length scale~$\int c_s d\t$. 
To preserve this scale in our model,  
where $\rho_{\g}$ is replaced by $\rho_{\rm rel}$,
we increase the baryon density by a factor
$(1+R_{\nu})$ over its physical value:
\be
\rho_b^{\rm(model)}=\lf(1+R_{\nu}\rt)\rho_b^{\rm(phys)}~.
\lb{bar_breakdown}
\ee

Finally, the total mean density of non-relativistic 
matter~$\rho_{\rm m}$, including baryons and CDM, is taken to
equal its physical value:
\be
\rho_{\rm m}^{\rm(model)} = \rho_{\rm m}^{\rm(phys)}~.
\lb{matter_breakdown}
\ee
For eqs.\rfs{bar_breakdown}{matter_breakdown} to hold, 
the mean density of CDM in our model must be reduced 
slightly compared with its physical value:
$\rho_c^{\rm(model)}= \rho_c^{\rm(phys)}-R_{\nu}\rho_b^{\rm(phys)}$. 
With these definitions, we arrive at a self-consistent two fluid model 
that preserves the important time scale of the radiation-matter density
equality as well as the acoustic length scale~$\int c_sd\t$.
From now on, we drop the superscript~``(model)''.

\subsection{Dynamical equations}

The dynamics of perturbations in the fluid model
is governed by the linearized Einstein and 
fluid equations.  We give these
equations in the conformal Newtonian gauge,\cts{MaBert,Mukh,Bert_COSMO_2000},
and then derive an equivalent but a more intuitive 
and easier to solve system of equations.

Metric perturbations are induced by perturbations in 
the energy-momentum tensor 
$T^{\mu}{}_{\nu}=\sum_a T_a^{\mu}{}_{\nu}$ where $a$ runs over 
all radiation or matter species in our model
($a=\g,b,c$).
For every species, $T_a^{\mu}{}_{\nu}$ can be parameterized
by an energy density enhancement~$\d_a$ and a velocity potential 
$\vp_a$, denoted by $W$ in\ct{Bert_COSMO_2000}, 
 \bs
 \lb{Tij} 
\be  
T_a^0{}_0&=&-\left(\rho_a+\d\rho_a\right)~,\quad   \d\rho_a=\rho_a\d_a~, 
\lb{T00}\\
T_a^0{}_i&=& (\rho_a+p_a)\,v_a{}_i~,\quad  v_a{}_i = -\Nb_i \vp_a~, 
\lb{T0i}\\
T_a^i{}_j&=& \d^i_j\,(p_a+\d p_a)~.
\lb{T3}
\ee
 \es
The stress~$T_a^i{}_j$ is isotropic for perfect fluids and the
pressure perturbations equal $\d p_{\g} = \frac13\,\rho_{\g}\d_{\g}$ and 
$\d p_c = \d p_b = 0$. 
Tight coupling between photons and baryons implies the equality of 
mean local velocities of the radiation and the baryon fluids, 
Sec.~\ref{sec_slk}. 
Assuming adiabatic initial conditions,
\be
\d_b=\frac34\,\d_{\g}\equiv \frac34\,\d_r \qquad\mbox{and}\qquad  
\vp_b=\vp_{\g}\equiv\vp_r~
\lb{dw_b}
\ee
before recombination.

The linearized Einstein equations for metric perturbations
$\phi$ and $\psi$ in eq.\rf{pertrw} are given by eqs.\rf{Einst_gen}
of the Appendix.  The last of eqs.\rf{Einst_gen} states that for
perfect fluids, when anisotropic stress is negligible,
$\psi=\phi$.  Then the remaining equations are
 \bs
 \lb{Einst_fl}
\be
\Nb^2\phi-3\,\frac{\dot a}{a}\left(\dot\phi+\frac{\dot a}{a}\,\phi\right)
  = 4\pi Ga^2\sum_a\d\rho_a\,,\quad
\lb{Einst1}
\\
\dot\phi+\frac{\dot a}{a}\,\phi = 4\pi Ga^2\sum_a(\rho_a+p_a)\vp_a\,,\quad
\lb{Einst2}
\\
\ddot\phi+3\,\frac{\dot a}{a}\,\dot\phi
 +\left[ 2\,\frac{\ddot a}{a} - \left(\frac{\dot a}{a}\right)^2 \right]\phi
  = 4\pi Ga^2\sum_a\d p_a\,,\quad
\lb{Einst3}
\ee
 \es
where ``$\bf \Nb$'' and ``$\dot{~}$'' 
denote partial derivatives with respect to the comoving
spatial coordinate~$\r$ and the conformal time~$\t$.

The fluid equations for density and velocity evolution in the conformal 
Newtonian gauge were derived in\ct{MaBert} and follow from the 
general formulae of Sec.~\ref{sec_gen}.  
In the tight coupling limit and our notations they are
 \bs
 \lb{fluid_eq}
\be
\dot\d_c&=& \Nb^2 \vp_c + 3\dot\phi~, \\
\dot \vp_c&=& -\,\frac{\dot a}{a}\,\vp_c + \phi~, \\
\dot\d_r&=& \frac43\,\Nb^2 \vp_r + 4\dot\phi~, 
\lb{fl_eq_dr}\\
\dot \vp_r&=& \fr1{1+(3\rho_b)/(4\rho_{\g})}\lf(\frac14\,\d_r - 
       \frac{3\rho_b}{4\rho_{\g}}\,\frac{\dot a}{a}\,\vp_r\rt)+\,\phi~.~~~~~
\lb{fl_eq_vr}
\ee
 \es
The scalar gravitational potential~$\phi$ does not present an independent
dynamical variable in addition to the densities and velocity potentials 
because on any given hypersurface of constant 
time the gravitational potential 
can be determined from the energy and momentum density perturbations on 
the same hypersurface via the generalized Poisson equation
\be
\Nb^2\phi = 4\pi Ga^2 \sum_a\left[
    \d\rho_a+3\frac{\,\dot a}{a}\,(\rho_a+p_a)\vp_a\right]~,
\lb{Pois_phi}
\ee 
following from eqs.\rfs{Einst1}{Einst2}.

We define 
\be
y(\t)\equiv \frac{a}{a_{\rm eq}}= \frac{\rho_{\rm m}}{\rho_{\rm rel}}~,\qquad
\beta\equiv \frac{\rho_b}{\rho_{\rm m}}~,
\lb{ybta_def}
\ee
where $a_{\rm eq}\simeq 1/(2.40\sh{\times}10^4\,\Omega_{\rm m} h^2)$ 
is the scale factor value at the time of
matter-radiation density equality
and $\rho_{\rm rel}=\rho_{\g}$ in our model. 
The ratio~$\beta$ is time independent.
An important scale in our problem is a 
characteristic conformal time of the transition from radiation to matter 
domination 
\be
\t_e\equiv \sqrt{\frac{a_{\rm eq}}{H_0^2\Omega_{\rm m}}}~.
\lb{te_def}
\ee
For the following $\Lambda$CDM model set of cosmological parameters:
$\Omega_{\rm m}=0.35$, $\Omega_{\Lambda}=0.65$, $\Omega_b h^2=0.02$, 
and $h\equiv H_0/(100{\rm \,km\,s^{-1}\,Mpc^{-1}})=0.65$, 
the numerical value of $\t_e$ is $c\t_e\simeq 130$\,Mpc.

With the above notations, the factor $4\pi Ga^2$ on the right hand
side of the Einstein equations becomes:
\be
4\pi Ga^2\rav \frac{3}{2\t_e^2}\,\frac1{y\rho_{\rm m}}~.
\lb{Ga2}
\ee
If the dark energy is neglected in the early epoch,
the Friedmann equation yields:
\be
\dot y^2 = \frac{1+y}{\t_e^2}~, \qquad
y(\t)=\frac{\t}{\t_e}+\frac14\left(\frac{\t}{\t_e}\right)^2~.
\lb{Frdmn_eq}
\ee 
The equality of matter and radiation 
densities, at which $y(\t_{\rm eq})\rav1$, occurs at 
$\t_{\rm eq}\rav 2(\sqrt 2\sh{-}1)\t_e\sh{\simeq}0.83\,\t_e$.

Dynamics of perturbations in our model depends on two characteristic
speeds, given by
\be
c_s^2=\frac{1}{3\lf(1+\fr34\,\b y\rt)}~
\lb{c_s_exp}
\ee
and
\be
c_w^2\equiv \left(\frac{\pd p}{\pd \rho}\right)_{\rm adiab}
=\frac{1}{3\lf(1+\fr34\,y\rt)}~,
\lb{c_w_def}
\ee
$p=\sum_a p_a$ and $\rho=\sum_a \rho_a$, see note~\cite{Note_on_c_w}.
Eq.\rf{c_s_exp} follows from the sound speed 
definition of eq.\rf{c_s_def}
since $\rho_b^{\rm(phys)}/\rho_{\g}^{\rm(phys)}
       =\rho_b^{\rm(model)}/\rho_{\g}^{\rm(model)}=\b y$.  
The second speed,
$c_w$, is not a true sound speed. It relates infinitesimal 
pressure and density changes for perturbations of constant radiation 
entropy per unit mass of non-relativistic matter,~$\eta$,
defined by
\be
\eta\equiv \d\left(\ln\frac{T^3_{r}}{\rho_{\rm m}}\right)
  = \frac{\sum\d p_a}{c_w^2\rho_{\rm m}}-\frac{\sum\d\rho_a}{\rho_{\rm m}}=
\lb{eta_def}\\
  = (1-\b)\left(\frac34\,\d_r-\d_c\right)~.
\nn
\ee
We describe~$\eta$ by a dimensionless
entropy potential~$\s$ such that
\be
\Nb^2\s\equiv \frac{3}{2\t_e^2}\,\eta~. 
\lb{sigma_def}
\ee
Using $\phi$, $\dot\phi$, $\s$, $\dot\s$ for independent
dynamical variables,
\be
  \ba{c}
\ddot\phi+\left(3+3c_w^2\right)\frac{\dot y}{y}\ \dot\phi
         +\frac{3c_w^2}{4\t_e^2y}\ \phi
    =c_w^2\,\Nb^2\left(\phi+\frac{\s}{y}\right)\,,\\
\ddot\s+\left(1+3c_w^2-3c_s^2\right)\frac{\dot y}{y}\ \dot\s
    =y\left(c_s^2-c_w^2\right)\Nb^2\left(\phi+\frac{\s}{y}\right)\,.
  \ea
\lb{phis_eq}
\ee
The first equation follows from the substitution of the
left hand side of eqs.~(\ref{Einst1},\,\ref{Einst3}) into the first line
of eq.\rf{eta_def} and using eqs.\rfd{sigma_def}{Frdmn_eq}.
The second is derived from 
eqs.~(\ref{sigma_def},\,\ref{eta_def},\,\ref{fluid_eq},\,\ref{Einst1},\,\ref{Einst2}).

Next, we replace the potentials~$\phi$ and $\s$ by
a pair of their linear combinations
$\phr\propto \phi+\s/y$ 
and~$\phc\propto \left(c_s^2/c_w^2-1\right)\phi-\s/y$,
which are chosen to diagonalize the second derivative terms in 
the system\rf{phis_eq}:
\be
  \ba{l}
\ddot\phr+\sum_{i=r,c}(a_{ri}\dot\phi_i+b_{ri}\phi_i)
                              =c_s^2\,\nabla^2\phr\ , \lb{phreq} \\
\ddot\phc+\sum_{i=r,c}(a_{ci}\dot\phi_i+b_{ci}\phi_i)=0~.
  \ea
\ee
The new variables $\phr$ and~$\phc$ are uniquely defined if
they are normalized so that
\be
\phi=\phr+\phc~.
\lb{phidec}
\ee
Then
\be
\s=y\left[\left(\fr{c_s^2}{c_w^2}-1\right)\phr-\phc\right]~
\lb{sdec}
\ee
and
 \bs
 \lb{phrc}
\be
\phr&=& \fr{c_w^2}{c_s^2}\left[\phi+\frac{\s}{y}\,\right],~\\
\phc&=& \left(1-\fr{c_w^2}{c_s^2}\right)\phi
     -\left(\fr{c_w^2}{c_s^2}\right)\frac{\s}{y}~.
\ee
 \es
The matrices $a_{ij}$ and $b_{ij}$ are obtained by
straightforward and somewhat tedious substitution of
eqs.~(\ref{phidec}--\ref{sdec}) in the system of 
eqs.~(\ref{phis_eq}).
The result is
\begin{widetext}
 \bs
 \lb{phrceq} 
\be
&&\ddot\phr+3\left(1+c_s^2\right)\frac{\dot y}{y}\,\dot\phi_r
             +3c_w^2\,\frac{\dot y}{y}\,\dot\phi_c
             +\,\frac{3\left(c_w^2+\varphi\right)}{4\t_e^2y}\,\phi_r
             -\frac{(2/c_s^2)-4+3c_w^2}{3\t_e^2y^2}\,\phi_c
               =c_s^2\,\nabla^2\phr\, , 
\lb{phrceq1}
\\
&&\ddot\phc-3\left(c_s^2-c_w^2\right)\frac{\dot y}{y}\,\dot\phi_r
             +3\,\frac{\dot y}{y}\,\dot\phi_c
             -\frac{3\varphi}{4\t_e^2y}\,\phi_r
             +\frac{(2/c_s^2)-3}{3\t_e^2y^2}\,\phi_c=0~
\lb{phrceq2}
\ee
 \es
\end{widetext}
where
$\varphi\equiv(1-\b)\lf[2+2c_s^2+12c_s^4
                            +3c_s^2c_w^2(2-3c_s^2-3c_w^2)\rt]$.

Eqs.\rf{phrceq} are our principal equations of perturbation
evolution in the model of two perfect fluids.
We would like to stress that these are causal hyperbolic
(wave) equations in contrast to ``action at a distance'' type
elliptic equation\rf{Pois_phi}.   This difference from the
original Einstein equations enables straightforward numerical
integration of eqs.\rf{phrceq} directly in real space. 

Eq.\rf{phrceq1} shows that perturbations 
in the radiation-baryon plasma, described by~$\phr$, 
propagate as acoustic waves with the sound speed~$c_s$.  
As the radiation wave passes by, its gravity perturbs locally
the cold dark matter as described by $\dot\phc$ and $\phc$
terms in eq.\rf{phrceq2}, and the CDM potential
$\phc$ starts evolving.

\subsection{Density and velocity fields}

In order to analyze the CMB temperature anisotropy or matter distribution 
in the universe, the potentials $\phr$ and $\phc$ should be related
to perturbations in density and velocity fields.  
The corresponding equations are derived in this subsection.

It is convenient to replace $c_w^2$ and $c_s^2$ by the following
variables linear in~$y$:
\be
A\equiv \frac1{3c_w^2}= 1+\frac34\,y~,\quad  
B\equiv \frac1{3c_s^2}= 1+\frac34\,\b y~.
\lb{AB_def}
\ee
For the quantities
\be
\d\equiv \frac{\sum\d\rho_a}{\rho_{\rm m}}~,\qquad
\vp\equiv \frac{\sum(\rho_a+p_a)\vp_a}{\rho_{\rm m}}~,
\lb{d&w_def}
\ee
the entropy perturbation $\eta$ of eq.\rf{eta_def}, 
and for an additional
variable~$\theta$ defined below, from 
eqs.~(\ref{Einst_fl},\,\ref{sigma_def},\,\ref{Ga2})
we find
\be
  \ba{l}
\d=\frac{B}{y}\,\d_r+(1-\b)\,\d_c
  =\frac{2\t_e^2}{3}\left[\Nb^2(y\phi)-3\,\frac{\dot y}{y}\,(y\phi)\dot{\,}\right],\\
\vp=\frac{4B}{3y}\,\vp_r+(1-\b)\,\vp_c
 =\frac{2\t_e^2}{3}\,(y\phi)\dot{\,}\,,\\
\eta=\vphantom{\frac{4B}{3y}}
     (1-\b)\left(\frac34\,\d_r-\d_c\right)
     =\frac{2\t_e^2}{3}\,\Nb^2\sigma~,\\
\theta\equiv \vphantom{\frac{4B}{3y}}
             (1-\b)\left(\vphantom{\frac34}\vp_r-\vp_c\right)                  
             =\frac{2\t_e^2}{3}\,\dot\sigma~.
  \ea
\lb{dw_def}
\ee
Since by the last two equations 
$\dot\eta+\bm{\Nb}\cdot{\bf j}_{\eta}=0$ with
${\bf j}_{\eta}\equiv-\bm{\Nb}\theta$,
physically, $\theta$ is a potential for the entropy 
current density.

The above equations become particularly 
symmetric in $(y\phi)$ and $\sigma$ if
the perturbation of energy density is described 
in terms of 
\be
\eps\equiv \d + 3\,\frac{\dot y}{y}\,\vp~.
\lb{eps_def}
\ee
Then by eqs.\rf{dw_def} 
\be
  \ba{l}
\eps= \frac{2\t_e^2}{3}\,\Nb^2(y\phi)~,
  \ea
\ee
and $\eps$ satisfies the conservation equation
$\dot\eps+\bm{\Nb}\cdot{\bf j}_{\eps}=0$ with 
${\bf j}_{\eps}=-\bm{\Nb}\vp$.

The linear combination on the right hand side of eq.\rf{eps_def}
remains invariant under
redefinition of constant-time hypersurfaces, $\~\t=\t+\alpha(\r,\t)$,
and spatial coordinates, $\~\r=\r+\bm{\Nb}\beta(\r,\t)$,
when in the linear order
\be
\~\d_a=\d_a-\frac{\dot\rho_a}{\rho_a}\,\alpha~,\qquad
\~\vp_a=\vp_a-\alpha~.
\lb{gauge_transf}
\ee
(For any $\alpha\sh{\not=}0$ or $\beta\sh{\not=}0$,
the metric~$ds^2$ will no longer be conformally isotropic in space
as it is in eq.\rf{pertrw}.) 
Using the equation of energy conservation,
\be
\dot\rho_a=-3\,\frac{\dot a}{a}\,(\rho_a+p_a)~,
\nn
\ee
we can see that 
\be
\~\eps=\eps~,\quad\~\eta=\eta~,\quad\~\theta=\theta~.
\nn
\ee    
Hence the invariant variable $\eps$ of eq.\rf{eps_def}
equals the energy density perturbation~$\~\d$ in the 
gauge where $\~\vp$ is identically zero, {\it i.e.}
the combined momentum density of both fluids vanishes.
Such a gauge is a generalization
of the comoving gauge to our multicomponent system,
and eq.\rf{eps_def} generalizes Bardeen's gauge 
invariant variable~$\eps$ introduced in\ct{Bard80}.

Using eqs.~(\ref{dw_def}--\ref{eps_def}) and
the definitions of $\phr$ and $\phc$ in 
eqs.~(\ref{phidec}--\ref{sdec}), 
it is straightforward to find:
 \bs
 \lb{dv_eqs}
\be
 \ba{c}
\d_r=\eps_r-\frac{3\dot y}{A}\,\vp~,\quad
\d_c=\eps_c-\frac34\,\frac{3\dot y}{A}\,\vp
 \ea~~
\lb{d_expl}
\ee 
where
\be
 \ba{l}
\eps_r\equiv \frac{y}{A}\left(\eps +\eta\right)
   =\frac{2\t_e^2}{3}\frac{y}{A}\,\Nb^2\left(\frac{A}{B}\,y\phr\right)~,\\
\eps_c\equiv \frac{3y}{4A}\,\eps -\frac{B}{A(1-\b)}\,\eta
      = \frac{2\t_e^2}{3}\,\frac{1}{1-\beta}\,\Nb^2(y\phc)~,
 \ea
\lb{e_expl}
\ee
and
\be
 \ba{l}
\vp_r=\frac{3y}{4A}\left(\vp+\theta\right)
   =\frac{2\t_e^2}{3}\,\frac{3y}{4A}\left(
           \frac{A}{B}\,y\phr\right)\dot{\vphantom{l}}~,\\
\vp_c=\frac{3y}{4A}\,\vp-\frac{B}{A(1-\b)}\,\theta
   =\frac{2\t_e^2}{3}\,\frac{1}{1-\beta}\left[
     \left(y\phc\right)\dot{\vphantom{a}}
  -\frac{(A/B)\dot{\vphantom{w}}}{(A/B)}\,y\phr\right].~\,
 \ea
\lb{w_expl}
\ee
 \es

In addition to determining the density and velocity perturbations,
eqs.\rf{dv_eqs} also help to establish the physical meaning
of the potentials $\phr$ and $\phc$.  From eqs.\rf{e_expl}, 
remembering eq.\rf{Ga2},
  \bs
\be
\Nb^2\phr&=& 4\pi Ga^2 \lf(\rho_{\g} \eps_r + \fr34\,\rho_b\eps_r\rt)~, 
\nn
\\
\Nb^2\phc&=& 4\pi Ga^2 \rho_c\eps_c~.
\nn
\ee
  \es
Thus, the potentials $\phr$ and $\phc$ are the metric
perturbations generated individually by the (comoving gauge)
disturbances in the radiation-baryon plasma and CDM.

The following conformal Newtonian gauge 
quantities contribute to the primary CMB temperature anisotropy
in the tight coupling regime,\cts{CMBFAST,Selj_fluid}: 
$\frac14\,\d_r\sh{+}\phi$, $\vp_r$, 
and~$\dot\phi$. Of these, $\vp_r$ is related to $\phr$
by the first of eqs.\rf{w_expl} and $\phi$ is given by eq.\rf{phidec}.
We can compute $\frac14\,\d_r\sh{+}\phi$ from
$\vp_r$ and~$\phi$ as 
\be
\frac14\,\d_r+\phi=\dot \vp_r+\frac34\,\beta\left[
         \lf(y\vp_r\rt)\dot{\vphantom{l}}-y\phi\right]~.
\lb{D_eff}
\ee
This equation follows from eq.\rf{fl_eq_vr}.

\section{Green's Functions and Initial Conditions}
\label{sec_inc} 

%
\subsection{Defining equations}

The evolution of cosmological perturbations in the linear
regime may be studied in position space by superimposing 
their individual Green's functions.
For the Green's functions one can take any complete
set of solutions of eqs.\rf{phrceq} satisfying
suitable initial conditions.
We consider two types of initial conditions: 
An initially point-like perturbation at the origin, {\it e.g.} 
\be
\phi^{(3)}(r,\t)\to \dd^{(3)}(\r)\quad\mbox{as}~~\t\to 0~
\lb{ic3}
\ee
where $\dd^{(3)}(\r)$ is the product of spatial coordinate delta 
functions $\dd(x)\dd(y)\dd(z)$, and a sheet-like perturbation that
is produced on the whole $(0,y,z)$ plane and is
independent of the $y$ and $z$ coordinates,\ct{BB_let}, {\it e.g.}
\be
\phi^{(1)}(x,\t)\to \dd(x)\quad\mbox{as}~~\t\to 0~.
\lb{ic1}
\ee
The initial conditions in the form\rf{ic1} will prove to be the most 
attractive for describing perturbation evolution.

All the fields $\phr$, $\phc$, $\d_r$, etc.\ are functions of 
three spatial coordinates $x$, $y$, and $z$, in addition to $\t$.  
If, however, a configuration of fields has no initial dependence on 
$y$ and $z$, as in eq.\rf{ic1}, the fields will remain independent of $y$ and $z$
for all time.  For such a configuration one needs to
solve the evolution equations\rf{phrceq} in the $x$ and $\t$ 
coordinates only.  Thus a Green's function for the evolution
equations with the plane-parallel initial conditions, eq.\rf{ic1}, 
is also the Green's function for the same equations considered in 
one spatial dimension.  
We denote these Green's functions
by the superscript~``$(1)$'',
and use the superscript~``$(3)$'' to distinguish the perturbations 
originating from a point disturbance, as in eq.\rf{ic3}.

When setting the initial conditions, one should also specify the ratio
of perturbation magnitude for different species, in our case
the radiation and the dark matter.  
In this paper we consider only adiabatic  
initial conditions, which are favored by the present 
experimental data,\cts{boom,DASI,maxima,toco,TegZal}, 
and by the simplest inflationary models,\cts{inf_models}:
\be
\fr{\sigma}{y\phi}\,\to\, 0\,\quad\mbox{as}\quad\t\to 0~.
\lb{adiab}
\ee  
(A Green's function approach to isocurvature initial conditions, arising 
naturally in topological defect models,\ct{defects}, has been considered
in\cts{VeerSteb90,Magueijo92}.)

Given the initial ratio of perturbations in various species, 
{\it e.g.\/} as implied by eq.\rf{adiab}, 
the subsequent evolution is completely described by
a single set of Green's functions, either one-dimensional or 
three-dimensional, for the relevant variables, in our case
$\phr$ and $\phc$. 
To illustrate the use of Green's functions, 
consider the total gravitational potential $\phi$.
The time dependence of its Fourier modes is given by
\be
\phi(\k,\t)= \phi^{(T)}(k,\t)\,A(\k)~. 
\lb{trf_def} 
\ee
Here, $A(\k)$ is a primordial potential perturbation created by 
inflation,\ct{Guth_infl}, or another mechanism of perturbation
generation, {\it e.g.\/}\cts{defects,ekpyrotic},
and $\phi^{(T)}(k,\t)$ is the Fourier space transfer function
\be
\phi^{(T)}(k,\t)= \int^{+\infty}_{-\infty}dx\,e^{-ikx}\,\phi^{(1)}(x,\t)~
\lb{phi1k}
\ee 
or
\be
\phi^{(T)}(k,\t)= \int d^3x\,e^{-i\k\cdot\r}\,\phi^{(3)}(r,\t)~.
\lb{phi3k}
\ee
Comparing eq.\rf{phi3k} and eq.\rf{phi1k}, one can relate the
two types of Green's functions as
\be
\phi^{(1)}(x,\t)=\int^{+\infty}_{|x|}2\pi\,r\,dr\,\phi^{(3)}(r,\t)~.
\lb{3to1}
\ee
Eq.\rf{3to1} says that in order to find the magnitude 
of the perturbation at a distance~$x$ from the plane of the initial 
excitation\rf{ic1}, one should add up the perturbations originating at all 
the points on the plane which are separated from the considered point
by $r=\sqrt{x^2+y^2+z^2}$. 
Conversely, a three-dimensional Green's function can always
be obtained from a plane-parallel one by differentiation: 
\be
\phi^{(3)}(r,\t)=-\,\frac{1}{2\pi r}\,\frac{\pd\phi^{(1)}(r,\t)}{\pd r}~.
\lb{1to3}
\ee

\subsection{Radiation era solutions}

When radiation dominates the background density, 
\be
y\simeq \fr{\t}{\t_e}\ll 1~,
\nn
\ee
$c_s^2\simeq 1/3$ and $c_s^2-c_w^2\simeq (1/4)(1-\b)\,y$.

For the adiabatic initial conditions\rf{adiab},
eqs.\rf{phrc} give $\phi_c\sh{\sim}y\phi_r$ during the radiation era.
Retaining in eqs.\rf{phrceq} the leading in $y$ terms only,
 \bs
\be
\ddot\phr+\frac4{\t}\,\dot\phr &=& \frac13\,\Nb^2\phr~, 
\nn
\\
\ddot\phc+\fr3{\t}\,\dot\phc+\fr1{\t^2}\,\phc 
&=& 3(1-\beta)y\lf[\fr1{4\t}\,\dot\phr+\fr1{\t^2}\,\phr\rt]\,.
\nn
\ee
 \es
Their non-singular one-dimensional Green's function solutions satisfying 
initial conditions\rf{ic1} and\rf{adiab} are
 \bs
 \lb{phrc_rad}
\be
\phr^{(1)}(x,\t)
    &=&\fr{3}{4}\,\fr{(c_s\t)^2-x^2}{(c_s\t)^3}~\theta(c_s\t-|x|)~,
\lb{phr_rad}\\
\phc^{(1)}(x,\t)
 &=&\fr{9(1\sh{-}\beta)y}{4}\lf[\fr{3(c_s\t)^2-2c_s\t|x|-x^2}{4(c_s\t)^3}
     \rt.-~\nn\\
 &\,&\qquad-\lf.\fr{|x|}{(c_s\t)^2}\ln\fr{c_s\t}{|x|}\right]\theta(c_s\t-|x|)~,
\quad\quad
\lb{phc_rad}
\ee 
 \es
where $\theta$ is the Heaviside step function: $\theta(x')=1$
for $x'>0$ and $\theta(x')=0$ otherwise.
The Fourier transforms of these Green's functions are the ``growing
mode'' transfer functions
\be
\phr^{(T)}(k,\t)&=&3\left(\frac{\sin z}{z^3} - \frac{\cos z}{z^2}\right)~, \qquad 
  \ba{c}
z\equiv kc_s\t~,
  \ea
\lb{trf_rad}
\\
\phc^{(T)}(k,\t)&=&\frac{9(1\sh{-}\beta)y}{2}\left[\frac{\sin(z) - z}{2z^3}
                    + \frac{\ln z + C - \ci(z)}{z^2} \right]
\nn
\ee 
where $C\rav 0.5772...$ is the Euler constant and 
\be
 \ci(z)\equiv -\int^{\infty}_{z} \frac{\cos z'}{z'}\,dz' =
            \ln z +C +\int^z_0\frac{\cos z' - 1}{z'}\,dz'
 \nn
\ee
is the cosine integral.

\section{Propagation of Perturbations}
\label{sec_res}

The one dimensional Green's functions for eqs.\rf{phrceq} at 
a later time can be obtained by numerical integration of these equations
starting from the radiation era solutions\rf{phrc_rad} 
at some $\t_{\rm init}\ll \t_e$.
The applied numerical methods are described and compared 
with Fourier space calculations in subsection~\ref{sub_num}.
The results of the integration up to the time of
hydrogen recombination at redshift $z\simeq 1.07\times 10^3$
are presented in Fig.~\ref{f_phrc}.
\begin{figure}[tb]
\includegraphics{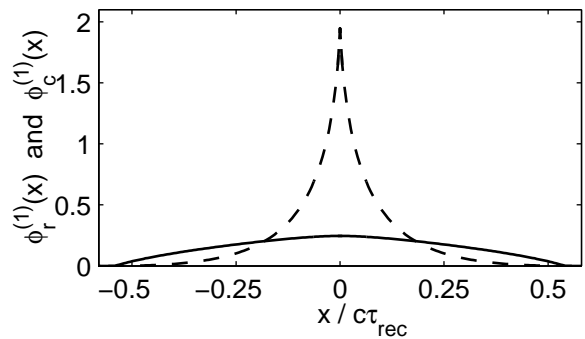}
\caption{
  Green's functions for the potentials~$\phr$ (solid) 
  and $\phc$ (dashed) at the time of recombination 
  $\t_{\rm rec}\simeq 2.15\,\t_e$ with 
  $\Omega_{\rm m}=0.35$, $\Omega_{\Lambda}=0.65$, 
  $\Omega_b h^2=0.02$, and $h=0.65$.
}  
\lb{f_phrc}
\end{figure}
The original delta function perturbation has separated into
left-going and right-going waves, whose evolution spreads
$\phr^{(1)}$ over space and diminishes the magnitude of 
the radiation disturbance. 
The gravitational potential hill of the radiation causes
outward-directed gravitational forces which expel the CDM away
from $x=0$.   Three snapshots of the time evolution of 
$\phr^{(1)}$ and $\phc^{(1)}$ 
are shown in Fig.~\ref{f_phis}.  Only the range
$x>0$ needs to be calculated since the potential Green's functions 
are even functions of~$x$.  
\begin{figure}[tb]
\centerline{\scalebox{0.8}[0.8]{\includegraphics{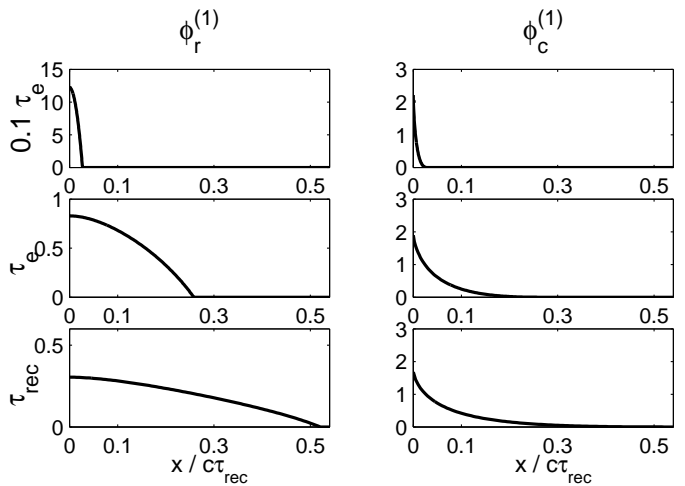}}}
\caption{
  Time evolution of the Green's functions of Fig.~\ref{f_phrc}
  describing the gravitational potentials for radiation (left) 
  and CDM (right). 
  The plots, from top to bottom, correspond 
  to the conformal time values $0.1\,\t_e$, $\t_e$, and 
  $\t_{\rm rec}\simeq 2.15\,\t_e$.
}
\lb{f_phis}
\end{figure}

The Green's functions are identically zero beyond the 
acoustic horizon, $|x|>S(\t)$, where
\be
S(\t)&\equiv&\int_0^{\t}c_s(\t')\,d\t'=
\lb{S_def}\\
   &=&\frac{4\t_e}{3\sqrt{\beta}}\ln\lf(
      \frac{\sqrt{1+\frac34y\beta}+\t_e\dot y\sqrt{\frac34 \beta}}
          {1+\sqrt{\frac34\beta}}\rt)\nn
\ee
($S(\t)\simeq \t/\sqrt{3}$ when $\t\ll\t_e$).
In a more careful treatment one will find weak precursors extending 
beyond the Green's functions acoustic fronts up to the particle 
horizon~$x=\pm c\t$.
The precursors arise because of partial photon,
and in a full treatment neutrino, free streaming with the speed 
of light.  Postponing detailed discussion of the photon diffusion 
until Sec.~\ref{sec_slk} and of the neutrino free streaming until
a separate paper, we continue the analysis of perturbation
dynamics in the perfect fluid model.

The density and velocity perturbations of the photon-baryon and 
CDM fluids are determined from the $\phr$ and $\phc$ potentials
by eqs.\rf{dv_eqs}.
The corresponding Green's functions are plotted in 
Figs.~\ref{f_dens} and~\ref{f_vel}.  Characteristic features
in the density and velocity Green's functions and their connection
to cosmological parameters are discussed in subsection~\ref{sub_res}.  
The delta function spikes at the acoustic wavefronts of
$\d_r^{(1)}$ and $v_r^{(1)}$ in the fluid model are fully 
calculable analytically and are considered in subsection~\ref{WFsing}.
\begin{figure}[tb]
\includegraphics{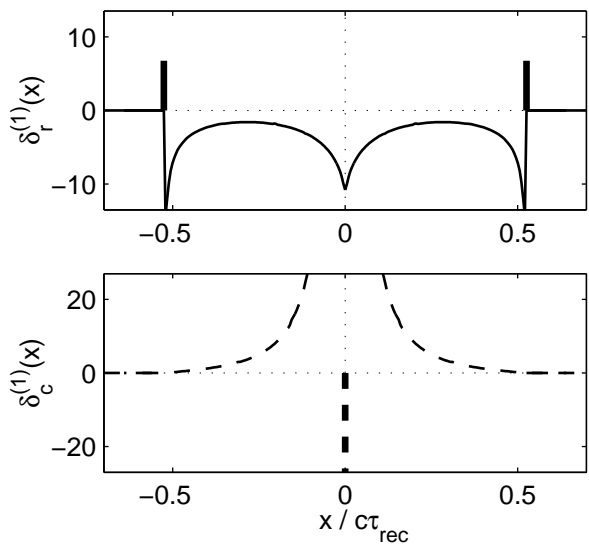}
\caption{Green's functions of the radiation density enhancement~$\d_r$
 (top) and the CDM density enhancement~$\d_c$ (bottom) at the 
 time of recombination.  The vertical spikes represent the delta-function 
 singularities of~$\d_r(x)$ at its wavefronts
 and of~$\d_c(x)$ at the origin.}
\lb{f_dens}
\end{figure}
\begin{figure*}[tb]
\includegraphics{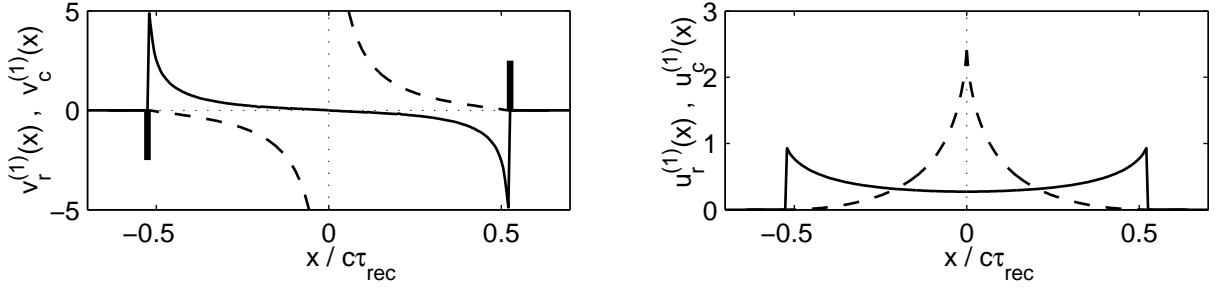}
\caption{Green's functions of the velocities 
$v_{r,c}\rav{-}\,\Nb \vp_{r,c}$ (left) and   
of the velocity potentials $\vp_{r,c}$ (right) 
at the time of recombination.
Solid and dashed lines are for the radiation and CDM respectively.}
\lb{f_vel}
\end{figure*}
\subsection{Sum rules}
\lb{SumRules}

Sum rules are simple, easy to apply integral relations that offer a powerful 
tool for checking analytical formulae or numerical calculations.  
The sum rules are more general than the fluid approximation:
they hold in the linear regime regardless of how complicated the 
internal dynamics of perturbations may be.  
They offer a simple method for estimating the accuracy 
of numerical calculations 
of the Green's functions.  In addition,
the sum rules give insight into the connection between
the position and Fourier space descriptions.  They become indispensable
for setting the initial conditions of Green's functions when one considers
perturbations to the full phase space distributions,\ct{SB_Thesis}.

The idea of the sum rules is to use eq.\rf{phi1k} with~$k=0$:
\be
\phi^{(T)}(k\,{=}\,0,\t)= \int^{+\infty}_{-\infty}dx\,\phi^{(1)}(x,\t)~.
\lb{k0mode}
\ee
The time evolution of the $k=0$ Fourier modes, the magnitude of which 
we denote by upper case letters, {\it e.g.} 
$\phi^{(T)}(k\,{=}0,\t)\equiv \Phi(\t)$, satisfies easily solvable
ordinary differential equations. The $k$-space equations at $k\rav0$ 
are especially simple because, given the adiabatic initial 
conditions, all the parameters characterizing relative motion of 
different matter and radiation components are unperturbed at
all time on infinitely large scales.  For example, in our case of the 
radiation-CDM model, 
\be
\Sigma(\t)\equiv \int dx\,\sigma^{(1)}(x,\t)= 0.
\lb{S_sum}
\ee

Trivially integrating first of eqs.\rf{phis_eq} over $x$ 
from $-\infty$ to~$\infty$, 
we obtain the equation for the gravitational 
potential in the $k=0$ mode: 
\be
\ddot\Phi+3\lf(1+c_w^2\rt)\frac{\dot y}{y}\,\dot\Phi
         +\frac{3c_w^2}{4\t_e^2y}\,\Phi=0~.
\lb{Phi_eq}
\ee
With $c_w$ given by eq.\rf{c_w_def},
the linear independent solutions of this equation are
\be
f_1&=&1+\frac{2}{9y}-\frac{8}{9y^2}-\frac{16}{9y^3}~,
\nn
\\
f_2&=&\frac{\sqrt{1+y}}{y^3}
 \simeq O(y)+\frac{1}{16}-\frac{1}{8y}+\frac{1}{2y^2}+\frac{1}{y^3}~.
\nn
\ee
Taking the linear combination of $f_1$ and $f_2$ satisfying
the initial condition~$\Phi(\t\to0)=1$, we obtain the sum rule 
for the gravitational potential:
\be
\Phi(\t)\equiv \int dx\,\phi^{(1)}(x,\t)
  =\frac{9}{10}\,f_1+\frac{16}{10}\,f_2~.
\lb{Phi_sum}
\ee
The sum rules for $\phr$ and $\phc$ then follow immediately
from eqs.(\ref{phrc},\ref{Phi_sum},\ref{S_sum}):
 \bs
\be
\Phi_r\equiv \int dx\,\phr^{(1)}(x,\t)&=&\frac{c_w^2}{c_s^2}\,\Phi~, 
\lb{Phr_sum}\\
\Phi_c\equiv \int dx\,\phc^{(1)}(x,\t)&=&\lf(1-\frac{c_w^2}{c_s^2}\rt)\Phi~.
\lb{Phc_sum}
\ee
 \es

The sum rules for the radiation velocity 
potential~$\vp_r$ and the combination~$\frac14\,\d_r\sh{+}\phi$
are relevant for CMB temperature anisotropy.  The first
sum rule simply follows from the first of eqs.\rf{w_expl} and 
from eq.\rf{Phr_sum}:
\be
U_r\equiv \int dx\,\vp_r^{(1)}(x,\t)
&=&\frac{3c_w^2\t_e^2y}{2}\lf(y\Phi\rt)\dot{\vphantom{l}}~.
\ee
The second can be derived starting from eq.\rf{D_eff} and
applying formulas of this subsection, including eq.\rf{Phi_eq},
\be
\frac14\,\D_r+\Phi&\equiv& \int dx\lf(
  \frac14\,\d_r^{(1)}+\phi^{(1)}\rt)=
\nn\\
&=&\dot U_r=-\,\frac{3c_w^2\t_e^2y^2}{2}\lf(
  \frac{\dot y}{y}\,\Phi\rt)\dot{\vphantom{\fr{w}{w}}}~.
\ee

Finally, we give the sum rules for the radiation and
CDM energy density enhancements.  They follow from 
eqs.\rfs{d_expl}{e_expl} and the second of eqs.\rf{dw_def}:
\be
\D_r=\fr43\,\D_c=-\,\frac{2\t_e^2\dot y}{A}\,\lf(y\Phi\rt)\dot{\vphantom{l}}~,
\ee
with $\D_r\equiv \int dx\,\d_r^{(1)}$ 
and $\D_c\equiv \int dx\,\d_c^{(1)}$. 

\subsection{Wavefront singularities}
\lb{WFsing}

We can obtain exact analytic formulae
describing Green's functions at the acoustic wavefronts 
$|x|=S(\t)$ in the fluid model.
As one will see below, the wavefront 
spikes in the density and velocity Green's functions 
play a significant role in the acoustic dynamics.
For the wavefront analysis, it is convenient to 
factor the radiation potential as
\be
\phr(x,\t)=C(\t)\,d(S(\t)-|x|,\t)
\lb{wf_ansatz}
\ee
where for the perfect fluids, with $x'\equiv S(\t)-|x|$,
\be
d(x'\le0,\t)\equiv0~,\quad \frac{\pd}{\pd x'}\,d(+0,\t)\equiv 1~,
\ee
{\it i.e.}\ $d(x',\t)\simeq x'\theta(x')$ as $x'\to 0$.
Substituting these equations in eq.\rf{phrceq1},
setting $|x|=S(\t)$, 
and taking into account that at the wave fronts
\be
\phr^{(1)}=\phc^{(1)}=0~,\quad \dot\phi_c^{(1)}=0~,
\ee
we find the following simple relation for $C(\t)$:
\be
2\dot C+\lf(4\,\fr{\dot y}{y}+3\,\fr{\dot c_s}{c_s}\rt)C=0~.
\lb{C_evol_eq}
\ee
Here we have applied the formula
\be
3c_s^2\,\fr{\dot y}{y}=\fr{\dot y}{y}+2\,\fr{\dot c_s}{c_s}
\nn
\ee
for the sound speed given by eq.\rf{c_s_exp}.
Integrating eq.\rf{C_evol_eq} and normalizing 
the result to agree with the radiation era 
solution~(\ref{phr_rad}) in the $y\to 0$ limit,
we obtain:
\be
C(\t)=\fr{9~~}{2\t_e^2y^2\lf(3c_s^2\rt)^{3/4}}~.
\lb{C_expl}
\ee
In particular, at $x=\pm S(\t)$,
\be
\dot\phr^{(1)}=\mp c_s\phr^{'(1)}
  =\frac{3\sqrt3~~}{2\t_e^2y^2\lf(3c_s^2\rt)^{1/4}}~.
\lb{phi_sing_sol}
\ee

The values of velocity potentials at the wave fronts 
follow from eqs.~(\ref{w_expl}):
\be
\vp^{(1)}_r=\frac{3\sqrt3}{4}\lf(3c_s^2\rt)^{3/4}~,~~~
\vp^{(1)}_c=0~.
\ee
Because $\phr$ and $\vp_r$ are identically zero 
when $|x|>S(\t)$, the velocity $v_r=-\Nb \vp_r$ and the
radiation density~$\d_r$, given by
eqs.~(\ref{d_expl}--\ref{e_expl}), both contain a delta function
singularity at the acoustic horizon:
 \bs
 \lb{sing}
\be
\d^{(1)}_{r,\rm~sing}
 =\eps^{(1)}_{r,\rm~sing}=3\lf(3c_s^2\rt)^{1/4}\dd(\,|x|-S)~,~~\,
\lb{dr_sing}
\\
v^{(1)}_{r,\rm~sing}
 =\frac{3\sqrt3}{4}\lf(3c_s^2\rt)^{3/4}\sign(x)\,\dd(\,|x|-S)~.~
\lb{vr_sing}
\ee
 \es

The dynamics of the shock-like singularities in the Green's functions 
is linear, even if nonlinear effects would 
take place in the propagation of a real shock wave in the photon-baryon 
plasma.  Indeed, the actual cosmological perturbations are the convolutions 
of the Green's functions with the smooth primordial potential field.  
As long as the overall perturbations are small,
they remain as regular as 
the primordial perturbation field and their dynamics is well in the 
linear regime. (On very small scales, there may exist physically 
relevant nonlinear effects,\ct{nonlin}, 
unrelated to the above Green's function singularities. These effects 
are absent when photons and baryons are tightly coupled.)

\subsection{Numerical integration}
\lb{sub_num}

From the hyperbolic (wave) nature of eqs.\rf{phrceq}, 
the values of the potentials
$\phr$ and~$\phc$ at a point $(x,\t+\D\t)$ are uniquely determined
by their past values within the interval $[x-c_s\D\t,x+c_s\D\t]$ 
at time~$\t$.
Cusps at the potential wavefronts will not be distorted by numerical
errors during the evolution if discretization intervals in space and 
time are related so that $c_s \D\t$ is a multiple of $\D x$.
In particular, for $c_s \D\t=\D x$, the second derivative terms in 
eqs.\rf{phrceq} may be approximated by the second order scheme 
\be
&&\ddot\phr(x,\t)-c_s^2\Nb^2\phr(x,\t)~\simeq 
\nn\\
&&\qquad
\simeq~ \fr{1}{(\D\t)^2}\,[\phr(x,\t+\D\t)+\phr(x,\t-\D\t)
\nn\\
&&\qquad\qquad~ 
-~\phr(x+c_s \D\t,\t)-\phr(x-c_s \D\t,\t)]~,\qquad 
\lb{fin_diff2}
\\
&&\ddot\phc(\t)~\simeq   
\nn\\
&&\qquad 
\simeq~\fr{1}{(\D\t)^2}\lf[\phc(\t+\D\t)
+\phc(\t-\D\t)-2\phc(\t)\rt]~. 
\nn
\ee
In practice, we use the horizon radius $S(\t)$ 
for the time evolution coordinate and
take $\D S=\D x$.

The first time derivatives of $\phr$ and $\phc$ can be approximated by 
an implicit second order scheme
\be
\dot\phi(\t)\simeq
     \fr{1}{\D\t}\lf[\phi(\t+\D\t)-\phi(\t)\rt]
    -\fr{\D\t}{2}\,\ddot\phi(\t)~.
\lb{fin_diff1}
\ee
An explicit scheme should be avoided as numerically unstable.
With the substitutions~(\ref{fin_diff2}--\ref{fin_diff1}), 
eqs.\rf{phrceq}
become finite difference equations that can be straightforwardly
solved for the values of 
$\phr$ and $\phc$ at~$(x,\t+\D\t)$.  The resulting
finite difference scheme is of the second order and it preserves 
the form of cusps or discontinuities in propagating waves. 

We start the integration from the radiation era solutions\rf{phc_rad}
at some small initial time, e.g. $\t_{\rm init}\simeq 10^{-4}\t_e$.
To follow the evolution over a large dynamic range,
the spacings of the space and time grids are increased by a factor of~$2$ 
every time the extent of the acoustic horizon~$S$ doubles.
Then the number of discretization points $N=S/\D x$ used to 
represent the Green's functions at all later times remains in the range
$N_{\rm min}\le N\le 2N_{\rm min}$ where $N_{\rm min}$ is the initial
number of points within the horizon.
The total number of $\phr$ and $\phc$ evaluations needed to
evolve the potentials to a final time~$\t_{\rm fin}$ is of the order 
of~$N^2\ln(\t_{\rm fin}/\t_{\rm init})$.

Whenever a Fourier space transfer function is needed,
it can be obtained from the corresponding Green's function using the
Fast Fourier Transform (FFT) algorithm.  The FFT can be efficiently
applied to Fourier integrals as described in\ct{NR} with its CPU 
time scaling as $N\log_2N$.
Due to the finite extent of the Green's functions, the related
transfer functions oscillate in $k$ and~$\t$
requiring in general a larger number of evaluations for 
their direct Fourier space calculation with the same accuracy.

The sum rules introduced in Subsection~\ref{SumRules}
prove to be a highly efficient debugging tool available for
calculations in position space.  The sum rule relations 
can also be used to estimate the numerical error of computations.

Calculations of the Green's functions in Figs.~\ref{f_phrc} and~\ref{f_phis}
proceeded from $\t_{\rm init}\simeq 10^{-4}\t_e$ to 
$\t_{\rm rec}\simeq 2.15\,\t_e$.  The corresponding CPU time 
for our Pentium IV PC is $0.04\,s$
with the minimal number of points in the spatial grid $N_{\rm min}=64$
and $0.1\,s$ with $N_{\rm min}=128$.  The accuracy of these calculations
as estimated by the sum rules is $0.19\%$ and $0.05\%$ respectively.

\subsection{Discussion of Green's function features}
\lb{sub_res}

The plane parallel Green's functions for the potentials 
$\phr$ and $\phc$, plotted in Figs.~\ref{f_phrc} and~\ref{f_phis},
are continuous functions of $x$ and $\t$.
In the fluid approximation, the radiation potential is 
characterized by discontinuous first derivatives
(kinks) at its acoustic wavefronts as described by 
eq.\rf{phi_sing_sol}.   The CDM potential has a central cusp reflecting the 
initial repulsive singularity in the gravitational potential
$\phr^{(1)}$; this cusp is preserved because the CDM particles
have no thermal motion.  In fact, once the photon energy density
becomes negligible in the matter dominated era, the potential
$\phc(x,\t)$ stops evolving, as may be seen from eqs.\rf{phrceq} 
in the limit $y\gg1$. 

The discontinuities in the potential derivatives give rise to 
delta function singularities in the corresponding density and
velocity transfer functions, following from 
eqs.~(\ref{dv_eqs}) and plotted for the time of
recombination in Figs.~\ref{f_dens} and~\ref{f_vel}.  
The singularities are visualized in the figures by the vertical spikes 
at $x=\pm S$ for $\d_r$ and $v_r$ and at $x=0$ for $\d_c$.

The wavefront singularities in $\d_r^{(1)}(x)$ and $v_r^{(1)}(x)$
are described analytically by eqs.\rf{sing}. Their significance 
in perturbation dynamics may be characterized by the delta function 
contributions to the sum rule integrals
for the corresponding Green's functions.  
For the radiation energy density, we find that 
in our $\Lambda$CDM model at recombination  
$\int dx\,\d^{(1)}_r{}_{\rm sing}/\int dx\,\d^{(1)}_r
\simeq 2.7/(-2.3)$.  Thus the delta function singularities 
play a major role in the perturbation dynamics even on the 
largest scales.  Their relative contribution to the CMB 
anisotropy even increases on smaller scales because the oscillation
amplitude of the delta function Fourier transform
does not fall off with~$k$, as opposed to the oscillations in
the non-singular term Fourier transforms.
We have shown in\ct{BB_let} that in real space the singularities give rise
to a localized feature in the CMB angular correlation function~$C(\theta)$. 

The delta function singularities acquire a finite width
when photon diffusion, or Silk
damping,\ct{Silk_damp}, is included into the analysis.
When the photon mean free path is small 
compared with the scales considered, Silk damping 
can be approximated as 
\be
\d^{(T)}_r(k) \simeq
e^{-k^2x_S^2}\,\d^{(T)}_{r~\rm fluid}(k)~, 
\lb{Silk_damp0}
\ee
Refs.~\cite{PeebYu70,Weinb_damp,Kaiser}.  
As found in\ct{Kaiser},
\be
x_S^2(\t) \simeq \fr16\int_0^{\t}\lf[1-\fr{14}{15}\lf(3c_s^2\rt)
                     +\lf(3c_s^2\rt)^2\rt]
        \fr{cd\t}{an_e\sigma_T}~
\lb{x_s0}
\ee
where $n_e$ is the proper number density of electrons and $\sigma_T$
is Thomson scattering cross section.
Silk damping broadens a delta function singularity 
in a position space Green's function to a Gaussian of finite 
width: 
\be
\dd(|x|-S)\to \fr1{2x_S\sqrt{\pi}}~\exp\lf[-\,\fr{\lf(|x|-S\rt)^2}{4x_S^2}\rt]~,
\lb{diff}
\ee
as obtained by applying the filtering\rf{Silk_damp0}
to the constant Fourier transform of a delta function. 
We rederive eqs.\rfs{Silk_damp0}{x_s0} within our 
position space formalism in Sec.~\ref{sec_slk}.

The CDM density perturbations, with the Green's function shown in the bottom
panel of Fig.~3, eventually seed the formation of galaxies. The central spike
in the figure arises from $\Nb^2\phc$ in the second of eqs.\rf{e_expl}.
It is negative because of the repulsive sign of the initial
gravitational potential peak, as shown in the top row of Fig.~\ref{f_phis}.
The spike is surrounded by positive tails because the CDM
pushed out from $x=0$ piles up into the region between $x=0$
and the acoustic wavefront.

\begin{figure}[tb]
\centerline{\includegraphics{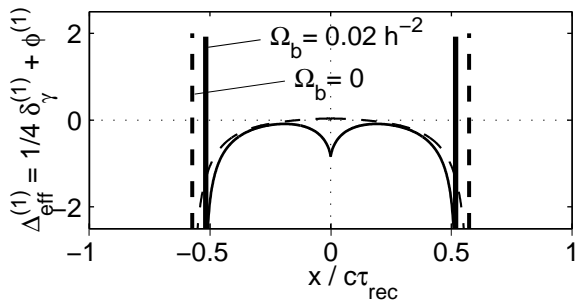}}
\caption{
 The position space transfer function for the intrinsic
 ($\frac14\delta_\gamma$) plus gravitational redshift contributions
 to the CMB temperature anisotropy $\D T/T$ at recombination
 assuming tight coupling between photons and baryons.
 The values of cosmological parameters are taken as
 $\Omega_{\rm m}=0.35$, $\Omega_{\Lambda}=0.65$, and $h=0.65$.
 Baryon drag effects are
 evident from comparing the solid, $\Omega_b h^2 = 0.02$,
 and the dashed, $\Omega_b = 0$, lines.
}
\lb{f_Deff0}
\end{figure}
To understand the physics of the dip in $\d_r^{(1)}(x)$ at $x=0$
in Fig.~\ref{f_dens},
let us examine the intrinsic radiation
temperature perturbation $\d_r/4$ corrected for the 
temperature redshift by the local metric perturbation at~$x$:
\be
\D_{\rm eff}\equiv \fr14\,\d_r+\phi~.
\lb{Deff_def}
\ee
The plane-parallel Green's function for $\Delta_{\rm eff}$ at
recombination is shown in Fig.~\ref{f_Deff0}.
The smoothness of
$\Delta_{\rm eff}^{(1)}$ at $x=0$ in the absence of baryons
(dashed line in Fig.~\ref{f_Deff0}) can be understood as the
condition of thermodynamic,\ct{Therm_eq}, and hydrostatic photon
equilibrium near $x=0$.  
(Obviously the equilibrium does not apply at the acoustic wavefronts, 
but it does hold as $x\to0$ where the elapsed time is many sound-crossing
times.)  Linearizing the relativistic equation of hydrostatic equilibrium,
\begin{equation}
   \nabla_i p+(\rho+p)\nabla_i\phi=0\ ,
\label{hseq}
\end{equation}
for the case of the radiation fluid containing no baryons
we obtain 
\begin{equation}\label{hseqpb}
   \nabla_i\lf({1\over3}\,\delta_\gamma+
     {4\over3}\,\phi\rt)=
     {4\over3}\,\nabla_i\Delta_{\rm eff}=0\ .
\end{equation}
Thus, if $\rho_b=0$, $\Delta_{\rm eff}$ has zero gradient in
equilibrium.  If $\Omega_b\ne0$, on the other hand,
the hydrostatic eq.\rf{hseq} applied to tightly coupled radiation-baryon 
fluid gives $\Nb_i\Delta_{\rm eff}=-(3/4)\b y\Nb_i\phi$.
The positive cusp of the CDM gravitational potential becomes a 
dip of CMB temperature.
The dip appears because the heavy pressureless
baryons are repelled by the cusp in the CDM gravitational
potential $\phc$ from the
$x=0$ region and they drag the coupled photon fluid away from the origin.
In eq.\rf{D_eff}, rewritten as
\be
\D_{\rm eff}=\dot\vp_r+\frac34\,\beta\lf(y\dot\vp_r+\dot y\vp_r-y\phr\rt)
                      -\frac34\,\beta y\phc~,
\lb{gully_quan}
\ee
only the last, $\phc$, term on the right hand side
has a cusp at $x=0$.  The contribution
of this term and thus the magnitude of the dip in~$\D_{\rm eff}$
are roughly proportional to
$\beta y(\t_{\rm rec})\propto \Omega_bh^2$.
Another effect of the baryonic constituent on the radiation
perturbations evident from Fig.~\ref{f_Deff0} is the decrease  
of the sound speed and thereby the reduction of the acoustic 
distance compared with a pure photon gas.

\section{Perturbation Dynamics Beyond the Fluid Model}
\label{sec_gen} 

Now we will distinguish between photon and baryon perturbations
and will not assume the fluid approximation for photons.
We will continue to exclude neutrino perturbations from
our model as described in Sec.~\ref{subsec_model}.

The linear evolution equations for
the photon and baryon velocity potentials 
follow from linearized Boltzmann equation and 
are given in the Appendix as the second of eqs.\rf{mult_gamma} 
and the second of eqs.\rf{mult_bar}.
In these equations, the terms proportional to the inverse of 
the mean conformal time of a photon free flight,
\be
\tt^{-1}\equiv {an_e\sigma_T}~,
\lb{tau_c_def}
\ee
are due to Thomson scattering between photons and baryons.
The Thomson scattering damps the velocity potential
difference, for which we define 
\be
\vp_d\equiv \vp_b-\vp_{\g}~.
\ee
However, the scattering does not affect the overall
momentum density of the plasma.
Defining a momentum-averaged velocity potential 
of photons and baryons 
\be
\vp_r\equiv \fr{(\rho_{\g}+p_{\g})\vp_{\g}+\rho_b \vp_b}
             {(\rho_{\g}+p_{\g})+\rho_b}
   = \fr1B\lf(\vp_{\g}+\fr{3\b y}{4}\,\vp_b\rt)~,
\ee
where $B$ is given by eq.\rf{AB_def},
from the velocity equations in eqs.~(\ref{mult_gamma},\,\ref{mult_bar})
we find:
\be
 \ba{l}
\dot \vp_r= \fr1B\lf(\fr14\,\d_{\g}+\Nb^2\pi_{\g}
         -\fr{3\b\dot y}{4}\,\vp_r\rt)+\phi~,\\
\dot \vp_d= -\fr{\dot a}{a}\lf(\vp_r+\fr1B\,\vp_d\rt)
          -\fr14\,\d_{\g}-\Nb^2\pi_{\g}
          -\fr1{\tt}\,\fr{4B}{3\b y}\,\vp_d~.   
 \ea
\lb{w_plasma_eq}
\ee

Substituting $\vp_{\g}=\vp_r-\fr{3\b y}{4B}\,\vp_d$
and $\vp_b=\fr1B\,\vp_d+\vp_r$ in the density evolution 
equations in eqs.~(\ref{mult_gamma},\,\ref{mult_bar}), and defining
\be
\d_d\equiv \d_b-\frac34\,\d_{\g}~,
\ee
we obtain
\be
  \ba{l}
\dot\d_{\g}=\fr43\,\Nb^2\lf(\vp_r-\fr{3\b y}{4B}\,\vp_d\rt)
                +4\dot\psi~,\\
\dot\d_d=\Nb^2\vp_d~.
  \ea 
\lb{d_plasma_eq}
\ee

The cold dark matter density and velocity
perturbations evolve as
\be
 \ba{l}
\dot\d_c= \Nb^2 \vp_c + 3\dot\psi~, \\
\dot \vp_c= -\frac{\dot a}{a}\,\vp_c + \phi~.
 \ea
\lb{CDM_eq}
\ee
The conformal Newtonian gauge potentials $\phi$ and $\psi$,
defined by eq.\rf{pertrw},
can be determined non-dynamically from the first, second,
and fourth relations of eqs.\rf{Einst_gen}:
 \bs
\be
\Nb^2\psi= \fr3{2\t_e^2y}
      \lf[\fr{B}{y}\,\d_{\g}+(1-\b)\d_c
          +\b\d_d+3\fr{\dot y}{y}\,\vp\rt]\,,~
\lb{Pois_sep}\\
\psi-\phi= \fr6{\t_e^2y^2}\,\pi_{\g} ~,
\qquad\qquad\qquad\qquad\qquad\qquad\quad
\lb{Panis_sep}
\ee
 \es
where formula\rf{Ga2} was applied. 
The variable~$\vp$ in eq.\rf{Pois_sep} 
is defined as previously by eq.\rf{d&w_def}.

Two independent variables are needed to describe
the species number density perturbations $\fr34\d_{\g}$,
$\d_b$, and~$\d_c$ relative to each other.  
For one such variable we use a potential generating the baryon 
density perturbation relatively to photons:
\be
\Nb^2\s_d\equiv \fr{3}{2\t_e^2}\,\d\lf(\ln\frac{\rho_b}{T^3_{\g}}\rt)
  = \fr{3}{2\t_e^2}\,\d_d~.
\lb{sbp_def}
\ee
The other can be the entropy potential~$\s$ from 
eqs.\rfd{sigma_def}{eta_def} that now equals
\be
\Nb^2\s&\equiv& \fr{3}{2\t_e^2}\,\d\lf(\ln\frac{T^3_{\g}}{\rho_{\rm m}}\rt)=
\nn\\
       &=&\fr{3}{2\t_e^2}\lf[(1-\b)\lf(\fr34\,\d_{\g}-\d_c\rt)
     -\b\d_d\rt].
\lb{s_def_sep}
\ee

The time derivatives of $\Nb^2\s_d$ or $\Nb^2\s$
equal a full~$\Nb^2$ of a certain linear combination
of velocity potentials~$\vp_a$.  
Differentiating both sides of eq.\rf{sbp_def} and eq.\rf{s_def_sep}
with respect to~$\t$, remembering eqs.\rfs{d_plasma_eq}{CDM_eq},
and lifting~$\Nb^2$, one obtains:
\be
 \ba{l}
\dot\s=\fr{3}{2\t_e^2}\lf[(1-\b)\lf(\vp_r-\vp_c\rt)
     -\fr{\b A}{B}\,\vp_d\rt]~,\\
\dot\s_d=\fr{3}{2\t_e^2}\,\vp_d~.
 \ea
\lb{dot_s_sep}
\ee

Taking another time derivative of both sides of eqs.\rf{dot_s_sep},
using the dynamical $\dot \vp_a$ equations from eqs.\rfd{w_plasma_eq}{CDM_eq}, 
and expressing all the resulting density and velocity perturbations 
in terms of $\Nb^2\psi$, as given by eq.\rf{Pois_sep},
$\Nb^2\s_a$, eqs.~(\ref{sbp_def},\,\ref{s_def_sep}),
and  $\dot\s_a$, eqs.\rf{dot_s_sep},
 \begin{widetext}
 \bs
 \lb{sphi_evol_sep}
\be
\ddot\s+\fr{\b c_s^2}{c_w^2}\,\ddot\s_d
 +\left(1+3c_w^2-3c_s^2\right)\fr{\dot y}{y}\,\dot\s
 +\fr{\b c_s^2}{c_w^2}\,\fr{\dot y}{y}\,\dot\s_d
    =y\left(c_s^2-c_w^2\right)\Nb^2\lf(\psi+\frac{\s}{y}
     +\fr{2\pi_{\g}}{c_w^2\t_e^2y^2}\rt)\,,
\lb{s1_evol_sep}
\\
\ddot\s_d+\lf(\fr{4}{9\b yc_s^2\tt}+\fr{\dot y}{y}\rt)\dot\s_d
 +\fr{9\dot y c_w^2}{4}\,\dot\s
    =-\,\fr{3c_w^2y^2}{4}\Nb^2\lf(\psi+\frac{\s}{y}
          +\fr{2\pi_{\g}}{c_w^2\t_e^2y^2}\rt)\,. 
\lb{s2_evol_sep}
\ee
Similarly to the fluid case, these equations are supplemented
by the formula following from the first line of
eq.\rf{eta_def}, eqs.\rfd{sigma_def}{Frdmn_eq},
and the Einstein equations\rf{Einst_gen}:
\be
\ddot\psi+\lf(3c_w^2+2\rt)\fr{\dot y}{y}\,\dot\psi
         +\frac{\dot y}{y}\,\dot\phi
         +\frac{3c_w^2}{4\t_e^2y}\,\phi
    =c_w^2\Nb^2\lf(\psi+\frac{\s}{y}
          +\fr{2\pi_{\g}}{c_w^2\t_e^2y^2}\rt)~.
\lb{phi_evol_sep}
\ee
 \es
 \end{widetext}
Here $\phi$ is expressed via $\psi$ and~$\pi_{\g}$ using
eq.\rf{Panis_sep}. 
The set of eqs.~(\ref{sphi_evol_sep},\,\ref{Panis_sep})
is not closed as long as $\pi_{\g}$
remains an independent variable. 

We split the total  potential~$\psi$
into potentials~$\psi_a$ generated by the individual species:
\be
\psi=\psi_{\g}+\psi_c+\psi_d~.
\lb{phi_sep}
\ee
The dynamics of the potentials~$\psi_a$ may be described
by coupled wave equations provided the decomposition\rf{phi_sep}
is performed according to the two following requirements: 
First, on small scales, when the velocity term in 
eq.\rf{Pois_sep} is negligible, every $\Nb^2\psi_a$ is proportional 
to the corresponding
$\d_a$ and is independent of the density of the other species.
Second, the potentials $\psi_a$ are given by linear combinations
of $\psi$ and the entropy potentials $\s$ and~$\s_d$.   
These linear combinations are unique and
are easily found to be
 \bs
 \lb{psi_a}
\be
\psi_{\g}&=& \fr{c_w^2}{c_s^2}\left(\psi+\frac{\s}{y}\,\right)~,\\
\psi_c&=& \left(1-\fr{c_w^2}{c_s^2}\right)\psi
       -\left(\frac{c_w^2}{c_s^2}\right)\frac{\s}{y}
       -\frac{\b\s_d}{y}~,\\
\psi_d&=& \frac{\b\s_d}{y}\vphantom{\fr{c_w^2}{c_s^2}}~.
\lb{psi_a_d}
\ee
 \es
The corresponding Laplacians are
 \bs
\be
\Nb^2\psi_{\g}&=&\fr3{2\t_e^2y}
      \lf(\fr{B}{y}\,\d_{\g}+\fr{B}{A}\,3\,\fr{\dot y}{y}\,\vp\rt),
\\
\Nb^2\psi_c&=&\fr3{2\t_e^2y}
      \lf[(1-\b)\d_c+
      \left(1-\frac{B}{A}\right)3\,\fr{\dot y}{y}\,\vp\rt],~~ 
\\
\Nb^2\psi_d&=&\fr3{2\t_e^2y}
      \lf(\b\d_d\vphantom{\fr{B}{A}}\rt).  
\ee
  \es

The evolution equations for the 
potentials $\psi_a$ are obtained by substituting 
eq.\rf{phi_sep} and
\be
\s=y\lf[\lf(\fr{c_s^2}{c_w^2}-1\rt)\psi_{\g}-\psi_c-\psi_d\rt],~~
\s_d=y\,\fr1{\b}\,\psi_d
\ee
in eqs.~(\ref{sphi_evol_sep},\,\ref{Panis_sep}). 
Upon the substitution, one finds a system of coupled
second order wave equations generalizing eqs.\rf{phrceq}.  
For simplicity, here we write these equations omitting
the terms with no 
or only first derivatives of the potentials 
$\psi_a$ and~$\pi_{\g}$ unless a term contains the large
damping parameter~$\tt^{-1}$.  The neglected terms contribute
only to the dynamics on large scales $\lambda\gtrsim \t$, 
when the fluid eqs.\rf{phrceq} can be used.  
With the remaining terms, one obtains that on scales
$\lambda\ll\t$
 \bs
 \lb{phi_a_evol_sep}
\be
\ddot\psi_{\g}+\ddot\psi_d &\approx& c_s^2\,\Nb^2\lf(\psi_{\g}
     +\fr{2\pi_{\g}}{c_s^2\t_e^2y^2}\rt)\,, 
\lb{phi_a_evol_sep1} \\
\ddot\psi_d+\fr{4}{9\b y c_s^2\tt}\,\dot\psi_d
   &\approx& -\,\fr{3\b y c_s^2}{4}\,\Nb^2\lf(\psi_{\g}
     +\fr{2\pi_{\g}}{c_s^2\t_e^2y^2}\rt),~~~~~\,
\\
\ddot\psi_c &\approx& 0~. 
\lb{phi_a_evol_sep3} 
\ee
 \es
In general, eqs.~(\ref{phi_a_evol_sep}) should be completed
by the evolution equations for the higher multipoles of the
radiation phase space density $f_{\g\,l\ge2}$ and polarization
multipoles $g_{\g\,l}$ that are given in the Appendix.
However, in the following section we find that in the next to the 
leading order in the photon-baryon coupling~$\tt$,
$\pi_{\g}\equiv \fr12\,f_{\g\,2}$
is fully determined by $\psi_{\g}$ and~$\dot\psi_{\g}$.  
Then, neglecting special features of 
neutrino dynamics, the above equations provide a closed system.

\section{Tight coupling: Next to the leading order}
\label{sec_slk} 

Now we can 
consider systematically how the general formulae of the previous section 
reduce to the fluid equations in the limit of tight photon-baryon coupling 
and find the $O(\tau_c)$ leading corrections to the fluid 
approximation.

First of all, $\dot \vp_d$ equation in eqs.\rf{w_plasma_eq}
and all the equations for $\dot f_{\g\,l\ge2}$ and $\dot g_{\g\,l}$ 
in the Appendix have the generic form
\be
\dot f_l = a - \frac1{r\tt}\lf(f_l-b\rt)~,
\lb{damped_evol}
\ee
where $r$ is a positive number and $a$ and $b$ are some 
linear combinations of variables other than $f_l$.
In the tight coupling regime, $\tt\ll\t$, the evolution
given by eq.\rf{damped_evol} will quickly, over a time of 
order $\tt$, drive
$f_l$ to $f_l\simeq b+r\tt(a-\dot b)$, as is evident
from the explicit solution of eq.\rf{damped_evol}
\be
 \ba{rcl}
f_l&=&b+\int^{\t}d\t'\,(a-\dot b)\,
   \exp\lf(-\int^{\t}_{\t'}\fr{d\t''}{r\tt}\rt)=\\
   &=&b+r\tt (a- \dot b) + O(\tt^2)~.
 \ea
\ee 
Thus at tight coupling the dynamical equations of the form\rf{damped_evol}
reduce to {\it algebraic} equations $f_l\simeq b$ up to~$O(\tt^0)$ 
or, if $b$ in eq.\rf{damped_evol}
is absent, $f_l\simeq r\tt a$ up to~$O(\tt)$.  

Applying this observation to the evolution of
the photon polarization averaged multipoles $f_{\g\,l}$ in 
eqs.\rf{mult_gamma} and to the polarization difference multipoles 
$g_{\g\,l}$ in eqs.\rf{mult_polar}, we see that each $l\ge3$ 
multipole is suppressed relative to the corresponding
lower multipole by an extra power of~$\tt$ .
Since $f_{\g\,1}=\fr43\,\vp_{\g}\sim \tt^0$,
we then find that $f_{\g}{}_l\sim \tt^{l-1}$ 
for $l\ge1$,
$\tilde g_{\g}{}_0\sim \tt$, $\tilde g_{\g}{}_1\sim \tt^2$,
and $g_{\g}{}_l\sim \tt^{l-1}$ 
for $l\ge2$. 
The equations for $\dot f_{\g\,2}$,
$\dot{\tilde g}_{\g\,0}$ and $\dot g_{\g\,2}$
reduce to the following algebraic relations:
\be
&&f_{\g\,2}\simeq \fr1{10}\lf(f_{\g\,2}+{\~ g}_{\g\,0}+g_{\g\,2}\rt) 
               +\tt\,\fr25\,f_{\g\,1}~,\qquad
\nn\\
&&{\~ g}_{\g\,0}\simeq \fr1{2}\lf(f_{\g\,2}+{\~ g}_{\g\,0}+g_{\g\,2}\rt)~,
\lb{alg_mult_rel}\\
&&g_{\g\,2}\simeq \fr1{10}\lf(f_{\g\,2}+{\~ g}_{\g\,0}+g_{\g\,2}\rt)~\nn
\ee
where the omitted terms are $O(\tt^2)$.
Similarly, from the second of eqs.\rf{w_plasma_eq},
\be
\vp_d\simeq -\tt\,\fr{3\b y}{4B}\lf(\fr14\,\d_{\g}
          +\fr{\dot y}{y}\,\vp_r\rt)~.
\lb{wbp_sol}
\ee 
Resolving the linear algebraic system\rf{alg_mult_rel} 
in terms of~$f_{\g\,1}$ and then remembering eq.\rf{wbp_sol}
we find: 
\be
\pi_{\g}\equiv \fr12\,f_{\g\,2}\simeq \tt\,\fr{4}{15}\,f_{\g\,1}\equiv
        \tt\,\fr{16}{45}\,\vp_{\g}\simeq
        \tt\,\fr{16}{45}\,\vp_r~.
\lb{pi_nlo}
\ee

Substitution of results~(\ref{wbp_sol},\,\ref{pi_nlo})
into the right hand side of $\dot\d_{\g}$
and~$\dot \vp_r$ equations in eqs.\rfd{d_plasma_eq}{w_plasma_eq} 
gives
\be
  \ba{l}
\dot\d_{\g}\simeq \fr43\lf[1+\tt\lf(\fr{3\b y}{4B}\rt)^2\fr{\dot y}{y}\rt]
            \Nb^2\vp_r
            +\fr13\,\tt\lf(\fr{3\b y}{4B}\rt)^2\Nb^2\d_{\g}
            +4\dot\psi~,\\
\dot \vp_r\simeq \fr1B \lf(\fr14\,\d_{\g}-\fr{3\b\dot y}{4}\,\vp_r
            +\tt\,\fr{16}{45}\,\Nb^2\vp_r\rt)+\phi~.
  \ea
\nn
\ee
Comparing these equations with their perfect fluid analogues,
eqs.~(\ref{fl_eq_dr}--\ref{fl_eq_vr}), one can identify 
two qualitatively new terms appearing in the  $O(\tt)$ order. 
The $\Nb^2\d_{\g}$ term in the first equation 
corresponds phenomenologically to heat conduction
and $\Nb^2\vp_r$ in the second to 
bulk viscosity of adiabatic scalar perturbations in the 
photon-baryon plasma.

If Thomson scattering did not partially polarize the scattered
radiation, in place of eqs.\rf{alg_mult_rel} we would find 
$f_{\g\,2}\simeq \tt\,\fr25\,f_{\g\,1}$.  This is $25\%$ less than 
the result\rf{pi_nlo} and would wrongly yield a 
$25\%$~smaller value of the radiation bulk viscosity.
The relevance of fluctuations in photon polarization
for dissipation of the perturbations in photon-baryon 
fluid was pointed out in\ct{Kaiser}.

The remaining undetermined quantity describing the relative
motion of photons and baryons is their number density difference~$\d_d$,
which enters the Poisson equation\rf{Pois_sep}.  We will calculate the
related potential $\s_d$ defined by eq.\rf{sbp_def}.
Within the $O(\tt)$ accuracy considered here
one can substitute the perfect fluid 
expressions~(\ref{dv_eqs}) into the right hand side
of eq.\rf{wbp_sol}.  Then replacing $\phr$ and~$\phc$
by $\phi$ and~$\s$ and using the second of eqs.\rf{phis_eq} to eliminate
$\Nb^2\lf(\phi+\s/y\rt)$, we find
\be
\vp_d\simeq 
  -\,\tt\,\fr{2\t_e^2}{3}\,\fr{3\b}{4(1-\b)}\lf(y\dot\s\rt)\dot{\vphantom{l}}~.
\ee 
The relation $\dot\s_d=3/(2\t_e^2)\vp_d$ from eq.\rf{dot_s_sep}
can then be integrated from~$\t=0$,
with $\lf.\s_d\rt|_{\t=0}=0$ for adiabatic perturbations, to a given~$\t$: 
\be
\s_d  \simeq -\,\tt\,\fr{3\b}{4(1-\b)}\,y\dot\s~.
\lb{s_d_nlo}
\ee

Thus to first order in~$\tt$ one can continue to
describe the perturbation dynamics by only two independent fields
$\phr\equiv\psi_{\g}$ and $\phc\equiv\psi_c$.  By 
eqs.~(\ref{psi_a},\,\ref{Panis_sep}), these fields reduce to 
our earlier definitions\rf{phrc} in $\tt\to 0$ limit.
In the $O(\tt)$ order, the potential $\psi_d$ is related to $\phr$ 
and $\phc$ via eq.\rf{psi_a_d} and eq.\rf{s_d_nlo},
in which $\s$ may be substituted by the $O(\tt^0)$ order
expression\rf{sdec}.

Since $\pi_{\g}$ in eq.\rf{pi_nlo} is an $O(\tt)$ quantity,
$\vp_r$~on the right hand side of eq.\rf{pi_nlo} 
may also be expressed in terms of
$\phr$ using the leading order relations\rf{w_expl}.  When the
corresponding expression for $\pi_{\g}$ and eq.\rf{s_d_nlo} are
used to close the dynamical equations of the previous subsection,
one finds two coupled differential equations for $\phr$ and $\phc$
that contain all the terms of eqs.\rf{phrceq} and additional 
terms proportional to~$\tt$. 
(Note that the third time derivative of $\s$ 
arising from $\ddot\s_d$ may be reduced within our $O(\tt)$ accuracy
by relations\rf{phis_eq} to terms containing a lesser 
number of time derivatives.) 
When $\tt\ll\t$, the $O(\tt)$ terms containing three derivatives
of the potential may still contribute appreciably to small
scale variations of the potentials when their higher $\t$ or 
spatial derivatives become increasingly large.
Retaining only the second derivative terms with $O(\tt^0)$
coefficients and all the third derivative terms, 
from eqs.~(\ref{phi_a_evol_sep1},\,\ref{phi_a_evol_sep3})
we arrive at the following system, valid on small scales
$\lambda\ll\t$:
 \bs
\be
\ddot\phr&\simeq& c_s^2\,\nabla^2\phr
        + 2\tt g\, \Nb^2\dot\phr\,,\qquad
\lb{nlo_evol_sm}
\\
\ddot\phc&\simeq& 0
\lb{nlo_evol_sm_cdm}
\ee
 \es
where
\be
g(\t)\equiv \fr16\lf(1-\fr{14}{15B}+\fr{1}{B^2}\rt)\,.
\lb{g_result}
\ee
The last term in eq.\rf{nlo_evol_sm} and so the expression\rf{g_result}
receive contributions from both $\ddot\psi_d$ and $\pi_{\g}$
terms in eq.\rf{phi_a_evol_sep1}.

The $\Nb^2\dot\phr$ term in eq.\rf{nlo_evol_sm} describes the famous 
Silk damping of perturbations in the photon-baryon plasma
on small scales,\ct{Silk_damp}.
In momentum space, the dispersion relation imposed by
eq.\rf{nlo_evol_sm} on a plane wave $\phr=A_r\exp{(i\k\cdot\r-i\omega\t)}$ is
\be
\omega^2+2i\tt g k^2\omega
        -k^2c_s^2=0~.
\ee
The solutions to first order in $\tau_c$ are
$\omega=\pm\,kc_s-i\g$ with the damping rate
\be
\g\simeq \tt g k^2~.
\ee
This rate coincides with the result of\ct{Kaiser}, 
where derivations were done with a different
approach that also took into
account the polarizing property of Thomson scattering.

One can quantitatively describe photon diffusion at the
sharp wavefront of $\phr^{(1)}(x,\t)$, as it evolves from 
$\t_{\rm init}\to 0$ to a certain finite $\t$, by considering the 
full eq.\rf{phrceq1} with the third derivative 
term $2\tt g\Nb^2\dot\phr$ added to its right hand side. 
We look for a solution in the wavefront region $\lf|S(\t)-|x|\rt|\ll\t$
using the ansatz\rf{wf_ansatz}:
\be
\phr^{(1)}=C(\t)\,d(x',\t)~,\quad
  x'\equiv S(\t)-|x|~
\nn
\ee 
where $C(\t)$ is given by eq.\rf{C_expl} and initially
$d(x',0)\approx x'\theta(x')$.  
Assuming that
\be
\lf|{\pd d}/{\pd x'}\rt|\gg \lf|d/\t\rt|~~
  \mbox{and}~~
\lf|{\pd d}/{\pd x'}\rt|\gg
   \lf|{\pd d}/{\pd \t}\rt|~
\lb{rm_lim}
\ee 
and remembering eq.\rf{C_evol_eq} as the condition
of cancellation of the ${\pd d}/{\pd x'}$ terms, we find:
\be
\fr{\pd^2d}{\pd x'\pd\t}\simeq 
  \tt g\,\fr{\pd^3d}{\pd x'{}^3}~.
\nn
\ee
Integrating over $dx'$ from $-\infty$ to a given~$x'$, we
arrive at the classical diffusion equation
\be
\fr{\pd d}{\pd\t}\simeq \tt g
\fr{\pd^2 d}{\pd x'{}^2}~.
\lb{diff_eq_derived}
\ee
The solution of this equation satisfies the second
condition in eq.\rf{rm_lim} provided $\tt$ is
much less than the characteristic scale over which
$d(x',\t)$~varies in~$x'$.
Individual Fourier modes of any function satisfying 
eq.\rf{diff_eq_derived} are damped as 
\be
d(k,\t)\simeq e^{-k^2 x_S^2(\t)} d(k,0)
\ee
where 
\be
x_S^2(\t)=\int_0^{\t}g(\t')\,\tt(\t')\,d\t'~,
\lb{x_s_res}
\ee
in agreement with eqs.~(\ref{Silk_damp0}--\ref{x_s0}).

By eq.\rf{nlo_evol_sm_cdm}, small scale CDM dynamics
is not affected by Silk damping.

\section{Conclusions}
\label{sec_con} 

We have shown how to reduce the linearized cosmological dynamics of 
gravitationally interacting species in the conformal Newtonian gauge to a 
system of coupled wave equations.  These equations indicate that the linear 
evolution of not only the density fluctuations but also of the corresponding 
gravitational potentials (metric perturbations) proceeds causally and 
locally, as opposed to the instantaneous action at a distance seemingly 
implied by the Poisson equation.  In the fluid approximation, 
a disturbance in the gravitational potential propagates through the 
photon-baryon fluid at the speed of sound.  The 
locality of the linear dynamics permits the efficient analysis of 
the perturbation evolution
using the Green's function method.  
The Green's functions are simply the Fourier transforms of 
the familiar transfer functions, but the causal nature of the Green's 
functions provides insights that were previously unknown.

We find that the Green's functions of primordial isentropic perturbations 
prior to photon decoupling are sharp-edged acoustic waves expanding with 
the speed of sound in the photon-baryon plasma.  As shown in Fig. 3, much 
of the integral weight of the photon density perturbation is localized at 
the acoustic wavefront of the corresponding Green's function.  When the 
finite mean free path to Thomson scattering is accounted for, the photon 
density perturbation is broadened to the width of the Silk damping 
length.  The photon dynamics in the presence of Thomson scattering is 
described by the Boltzmann equation, which we have analyzed to first order 
in the photon mean free path.  The result is a diffusive damping 
corrections to the radiation wave equation.  
We show how photon polarization affects 
Silk damping and provide an alternative derivation of the results of 
Kaiser,\ct{Kaiser}.

Another insight obtained with the Green's function method concerns the 
effects of baryons.  The baryonic component of the plasma is responsible 
for a distinctive central dip in the Green's function for the 
gravitationally redshifted CMB photon temperature perturbation 
$\frac{1}{4}\delta_r+\phi$ as shown in Fig. 5.  
This feature gives rise,\ct{SB_Thesis}, to the known effect of odd peak 
enhancement and even peak suppression,\ct{HuSug96}, 
in the CMB power spectrum.
It is readily understood in our analysis through the 
gravitational effect of the cold dark matter acting on the baryons.

In this paper we considered perturbations only in photons, baryons, and 
CDM.  The Green's function method can be applied to the linearized phase 
space dynamics of an arbitrary number of species with rather general 
internal dynamics and mutual interactions, including neutrinos,\ct{forw}.  
The advantage of the 
position space approach over the traditional Fourier space expansion is its 
simple and explicit treatment of acoustic and transfer phenomena underlying 
the dynamics of all particle species.  In addition to providing an 
intuitive, compact framework for the many effects shaping the 
fluctuations of matter and radiation, this approach can give simpler 
analytical and faster computational methods for the still challenging 
aspects of cosmological perturbations, such as those of relic 
neutrinos.  Application of the similar wave equations and the Green's 
function approach to other areas of astrophysics where gravity 
influences acoustic or transfer phenomena might be fruitful.

\appendix
\section{Boltzmann hierarchy
         in~position~space}

Two independent physical potentials are needed to specify 
scalar metric perturbations in the general case, see 
{\it e.g.\/} any of\cts{Bard80,KS84,Mukh,LiddleLyth93,MaBert}.
In the conformal Newtonian gauge, these are $\phi(\r,\t)$ and $\psi(\r,\t)$ 
defined by eq.\rf{pertrw}.
The linearized Einstein equations in a flat universe
reduce to the following equations,
following correspondingly from the $0$-$0$, $0$-$i$, summed~$i$-$i$,
and traceless~$i$-$j$ components of $G^{\mu}{}_{\nu}=
8\pi GT^{\mu}{}_{\nu}$,\ct{Bert_LesHouches}:
\begin{widetext}
\be
 \ba{rcl}
\Nb^2\psi-3\,\fr{\dot a}{a}\lf(\dot\psi+\fr{\dot a}{a}\,\phi\rt)
 &=& 4\pi Ga^2\sum_a \rho_a\d_a~,\\
\dot\psi+\fr{\dot a}{a}\,\phi &=& 4\pi Ga^2\sum_a(\rho_a+p_a)\vp_a~,
 \qquad \\
\ddot\psi+\fr{\dot a}{a}\lf(2\dot\psi+\dot\phi\rt)
 +\lf[ 2\,\fr{\ddot a}{a} - \lf(\fr{\dot a}{a}\rt)^2 \rt]\phi
 - \fr13\,\Nb^2(\psi-\phi) &=& 4\pi Ga^2\sum_a\d p_a~,\\
\fr13\,(\psi-\phi) &=& 4\pi Ga^2\sum_a(\rho_a+p_a)\pi_a~.
 \ea
\lb{Einst_gen}
\ee
\end{widetext}
The energy density enhancement~$\d_a$ and the velocity potential~$\vp_a$ 
of each of the 
matter or radiation species ``$a$'' are defined, as previously, in 
terms of the corresponding energy-momentum tensor components by eq.\rf{Tij}.
The variables $\d p_a$ and $\pi_a$, giving
the isotropic and anisotropic components of stress perturbation,
are defined by
\be
T_a^i{}_j&=& \d^i_j\,(p_a+\d p_a)+
\nn
\\
&&+~(\rho_a+p_a)\,\fr32\lf(\Nb^i\Nb_j-\fr13\,\d^i_j\Nb^2\rt)\pi_a~\,~~
\lb{pi_def}
\ee
where $\Nb^i=\Nb_i$, assuming negligible $3$-space curvature.
The anisotropic stress potentials $\pi_a$ vanish 
for perfect fluids and we see from the last of eqs.\rf{Einst_gen} that the gravitational
potentials $\phi$ and $\psi$ are then equal.

Six variables specify the coordinates of a particle
in phase space at a given time.  For them, we take the comoving  
coordinates of the particle $r^i$ and the comoving momenta
\be
q_i\equiv a p_i
\ee
where $p_i$ are the proper momenta measured by a comoving
observer,\cts{BondSzalay83,MaBert}.  The momentum coordinates~$q_i$
are {\it not} canonically conjugate to the variables~$r^i$;
the canonical momenta of a particle of mass~$m_a$ are
$P_i = m_a{dx_i}/{\sqrt{-ds^2}} = (1-\psi)q_i$.

The particle density in phase space is specified by the canonical
phase space distribution~$f_a(r^i,P_j,\t)$:
\be
dN_a=f_a(r^i,P_j,\t) d^3r^i d^3P_j~
\lb{f_def}
\ee
for every species of particles and their states of polarization~$a$.
Time evolution of the phase space distributions
is given by Boltzmann equation
\be
\fr{\pd f_a}{\pd\t}+\dot r^i\,\fr{\pd f_a}{\pd r^i}
+\dot q\,\fr{\pd f_a}{\pd q}+\dot n_i\fr{\pd f_a}{\pd n_i}
=\lf(\fr{\pd f_a}{\pd\t}\rt)_C~,
\lb{Bolz_gen}
\ee
where $f_a$ is considered as a function of the coordinates 
$r^i$, $q\sh{\equiv}|q_i|$, $n_i\equiv {q_i}/{q}$, and~$\t$.
The energy-momentum tensor of the species~$a$ is given in 
the conformal Newtonian 
gauge by the following simple expression up to the first order 
of cosmological perturbation
theory,\cts{MaBert,BondSzalay83}:
\be
T^{\mu}_a{}_{\nu}=\int d^3p_i\,\fr{p^{\mu}p_{\nu}}{p^0}\,f_a~
\lb{Tmunu_expl}
\ee
with $p^0\equiv -p_0 \equiv \sqrt{(q/a)^2+m_a^2}$ and
$p^i\equiv p_i = q_i/a$. We later omit the species label 
$a$ if referring to any sort of particles in general.

As in the tight-coupling case, an arbitrary perturbation
$\d f(\r,q,\n,\t)$ of the phase space distribution 
about the unperturbed distribution $f_0(q,\t)$ can be expanded 
over plane waves, in which the value of~$\d f$ at 
a given $q$, $\n$, and $\t$ remains constant along two spatial 
directions.   To describe the dynamics of such a plane wave, 
we set the spatial coordinates $y$ and $z$ to
vary along these invariant directions, and the coordinate~$x$  
to vary along the remaining direction.
The perturbation~$\d f(x,q,\n,\t)$ can be expanded over 
Green's functions  satisfying the delta function initial
conditions $\d f^{(1)}(x,q,\n,\t)\to A(q,\n)\,\d_{\rm D}(x-x_0)$ 
as $\t\to0$ with various initial locations~$x_0$ and functional
forms of~$A(q,\n)$.
In the translationally invariant background,
the time evolution of the Green's functions is independent of~$x_0$.

If $\varphi$ is the angle in the $y$--$z$ plane 
between the vector~$(n_y,n_z)$ and the axis~$y$,
we may expand an arbitrary $A(q,\n)$ in the Fourier series
\be
A(q,n_x,n_y,n_z)=\sum_{m=-\infty}^{+\infty}e^{im\varphi}\,A_m(q,n_x)~.
\ee
By definition, the scalar perturbations have initially zero $A_m$
for $m\not=0$.  For a homogeneous and isotropic background,
subsequent evolution according to the Boltzmann 
equation\rf{Bolz_gen} in the linear regime can not excite $m\not=0$ 
components of~$\d f$ through scalar perturbations alone.  
Therefore, we will consider only the perturbations 
that are axially symmetric about $x$ axis: $\d f=\d f(x,q,n_x,\t)$.

The number of coordinates needed to describe $\d f$
can be reduced further when the particle 
mass is zero and the relevant scattering cross-sections are
energy independent.
Then one can work with the energy averaged perturbation of 
the distribution,\ct{Lindquist},
\be
F(x,n_x,\t)\equiv
  \fr{\int q^2 dq\,q\,\d f(x,q,n_x,\t)}{\int q^2 dq\,q\,f_0(q)}\,=
\lb{Fdef}\\
  =\sum_{l=0}^{\infty}(2l+1)\,F_l(x,\t)\,P_l(n_x)~,
\nn
\ee
where $P_l$ is a Legendre polynomial.
The energy density enhancement, mean velocity, and anisotropic pressure
of the particles described by
an energy averaged distribution~$F$ are given by 
\be
 \ba{rcl}
\d&=&\int_{-1}^{1}\fr{dn_x}{2}\,F(n_x) = F_0~,\\
-\Nb \vp&=&\fr34\int_{-1}^{1}\fr{dn_x}{2}~n_x F(n_x)
 =\fr34\,F_1~,\\
\Nb^2\pi&=&\fr34\int_{-1}^{1}\fr{dn_x}{2}\lf(n_x^2-\fr13\rt)F(n_x)
 =\fr12\,F_2~
 \ea
\lb{dw_from_F}
\ee 
for every massless species~$a$.
In these and the following equations
$\Nb\equiv \fr{\pd}{\pd x}$,
since we are 
considering plane-parallel perturbations that are constant in the $y$ and 
$z$ directions.
Eqs.\rf{dw_from_F} follow directly from the 
definitions~(\ref{T00},\,\ref{T0i},\,\ref{pi_def})
and the formula\rf{Tmunu_expl}.

Continuing  our practice of reducing
the number of gradients
in the evolution equations, we define for all~$l$
\be
F_l\equiv(-1)^l\,\Nb^lf_l~.
\lb{fl_def}
\ee
The evolution equations for the variables
in eqs.~(\ref{dw_from_F}--\ref{fl_def})
can be derived in position space from Boltzmann 
equation\rf{Bolz_gen},\ct{SB_Thesis}, or written 
down by replacing momenta $k$ by spatial 
gradients in previously derived Fourier space equations, 
{\it e.g.}\ct{MaBert}.
We give the corresponding hierarchy equations in
the conformal Newtonian gauge\rf{pertrw} for 
neutrinos, photons, baryons, and CDM
particles.

The evolution of neutrinos is given by
\be
\dot\d_{\nu}&=&\fr43\,\Nb^2\vp_{\nu}+4\dot\psi~,
\nn\\
\dot \vp_{\nu}&=& \fr14\,\d_{\nu}+\Nb^2\pi_{\nu}+\phi~,
\lb{mult_nu}\\
\dot f_{\nu}{}_l&=& \fr{l}{2l+1}f_{\nu}{}_{(l-1)}+\fr{l+1}{2l+1}\,\Nb^2f_{\nu}{}_{(l+1)}~,
~~ l\ge2~,
\nn
\ee
with $\pi_{\nu}=\frac12\,f_{\nu}{}_2$ and $\vp_{\nu}=\frac34\,f_{\nu}{}_1$.

The collision terms for Thomson scattering of photons
and baryons were derived in Refs.~\cite{BE8487}.
The resulting equations read
\be
\dot\d_{\g}&=&\fr43\,\Nb^2\vp_{\g}+4\dot\psi~,
\nn\\
\dot \vp_{\g}&=&\fr14\,\d_{\g}+\Nb^2\pi_{\g}+\phi-\frac1{\tt}\lf(\vp_{\g}-\vp_b\rt)~,
\lb{mult_gamma}\\
\dot f_{\g}{}_l&=& \fr{l}{2l+1}\,f_{\g}{}_{(l-1)}
                    +\fr{l+1}{2l+1}\,\Nb^2f_{\g}{}_{(l+1)}\,- \nn\\
               && -\,\frac1{\tt}\lf\{f_{\g}{}_l-\fr{\d_{l2}}{10}
          \lf(f_{\g}{}_2+\~ g_{\g}{}_0+g_{\g}{}_2\rt)\rt\}\,,~~l\ge2~.
\nn
\ee
Again,  $\pi_{\g}=\frac12\,f_{\g}{}_2$, $\vp_{\g}=\frac34\,f_{\g}{}_1$,
and $\tt$ is defined by eq.\rf{tau_c_def}.
The quantities $g_{\g}{}_l(x,\t)$ and~$\~ g_{\g}{}_l(x,\t)$ 
are defined as
\be
  \ba{l}
G_{\g}{}_0= \Nb^2 \~ g_{\g}{}_0~,\\
G_{\g}{}_1= -\Nb^3 \~ g_{\g}{}_1~,\\
G_{\g}{}_l=(-1)^l\,\Nb^l g_{\g}{}_l~,\quad l\ge2  \ea
\lb{gl_def}
\ee
where $G$ is the energy averaged distribution describing the
difference of two linear photon polarization components.
The equations of their evolution are
\be
\dot{\~ g}_{\g}{}_0&=&
    \Nb^2 \~ g_{\g}{}_1 -\frac1{\tt}\lf\{ \~ g_{\g}{}_0 
  - \fr12\lf(f_{\g}{}_2+\~ g_{\g}{}_0+g_{\g}{}_2\rt)\rt\}\,,
\nn\\
\dot{\~ g}_{\g}{}_1&=& \fr{1}{3}\,\~ g_{\g}{}_0
                    +\fr{2}{3}\,g_{\g}{}_2
                -\frac1{\tt}\,\~ g_{\g}{}_1~,
\lb{mult_polar}\\
\dot g_{\g}{}_l&=& \fr{l}{2l+1}\,g_{\g}{}_{(l-1)}
                    +\fr{l+1}{2l+1}\,\Nb^2g_{\g}{}_{(l+1)}\,-\nn\\
               && -\,\frac1{\tt}\lf\{g_{\g}{}_l - \fr{\d_{l2}}{10}
          \lf(f_{\g}{}_2+\~ g_{\g}{}_0+g_{\g}{}_2\rt)\rt\}\,,
 ~~ l\ge2~,
\nn
\ee
where $g_{\g}{}_1\equiv \Nb^2 {\~ g}_{\g}{}_1$ in the last
equation at $l=2$.

Applying momentum conservation to photon-baryon interactions,
one can easily modify the non-relativistic fluid equations of baryon 
evolution to include Thomson scattering:
\be
\dot\d_b&=&\Nb^2\vp_b+3\dot\psi~,
\lb{mult_bar}\\
\dot \vp_b&=&-\fr{\dot a}{a}\,\vp_b+\phi-\fr1{\tt}\fr{4}{3\beta y}\lf(\vp_b-\vp_{\g}\rt)~.
\nn
\ee
The parameters $y$ and $\beta$ are defined by 
eq.\rf{ybta_def}.
Aside from the distinction between $\phi$ and $\psi$ metric perturbations, 
the equations of CDM evolution are the same as in the fluid model:
\be
\dot\d_c&=& \Nb^2 \vp_c + 3\dot\psi~, 
\\
 \dot \vp_c&=& -\frac{\dot a}{a}\,\vp_c + \phi~.
\nn
\ee

\begin{acknowledgments}

This work has benefitted from discussions with 
J.~R.~Bond, A.~Loeb, L.~Page, P.~J.~E.~Peebles, P.~L.~Schechter,
and U.~Seljak. We also thank A.~Shirokov for his help with 
numerical calculations.
Support was provided in part by NSF grant ACI-9619019, 
by Princeton University Dicke Fellowship,
and by NASA grant NAG5-8084.

\end{acknowledgments}

\bibliography{fl}

\begin{thebibliography}{37}
\bibitem[{Not()}]{Note_on_c_w}
\bibinfo{note}{Our choice of the subscript label for the ``isentropic sound
  speed'' $c_w$ is motivated by the conventional notation $w\equiv p/\rho$. The
  above variables are related as $c_w^2=w+\lf(\pd w/\pd \ln\rho\rt)_{\rm
  adiab}$.}
\end{thebibliography}

\end{document}